\documentclass{article}
\usepackage{jcappub}
\pdfoutput=1

\usepackage{amsmath,amssymb,bm,color,float,graphicx,mathrsfs}
\usepackage{Macro}

\def\hs{\widehat{s}}
\def\hd{\widehat{d}}

\def\hx{\widehat{x}}

\def\hatn{\widehat{n}}
\def\bx{\bm{x}}
\def\br{\bm{r}}
\def\bd{\bm{d}}

\def\bs{\bm{s}}

\def\dsrc{\delta^{\rm int}_{\rm s}}
\def\ncsrc{n_{\rm s}}
\def\npsrc{n_{\rm p}}
\def\wte{\epsilon_r}

\title{Analyzing clustering of astrophysical gravitational-wave sources: Luminosity-distance space distortions}

\author[a]{Toshiya Namikawa}

\affiliation[a]{Department of Applied Mathematics and Theoretical Physics, University of Cambridge, Wilberforce Road, Cambridge CB3 0WA, Unite Kingdom}

\emailAdd{tn334@cam.ac.uk}


\abstract{
We present a formulation of observed number density fluctuations of gravitational-wave (GW) sources in a three dimensional space. In GW observations, redshift identification for each GW source is a challenging issue, in particular, for high redshift sources. The use of observed luminosity distance as a distance indicator will be a simple yet optimal way for measuring the clustering signal. We derive the density fluctuations of GW sources estimated from observed luminosity distance and sky position of each source. The density fluctuations are distorted as similar to the so-called redshift space distortions in galaxy surveys but with several differences. We then show the two-point correlation function and multipole power spectrum in the presence of the distortion effect. We find that the line-of-sight derivative of the lensing convergence, which does not appear in the redshift-space distortions, leads to significant distortions in the observed correlation function. In addition, the lensing effect affects higher-order multipole power spectra and its signal-to-noise at high redshifts. 
}

\begin{document}

\maketitle
\flushbottom
% Contents %

%//////////////////////////////////////////////////////////////////////////////////////////%
\section{Introduction} \label{sec:intro}
%//////////////////////////////////////////////////////////////////////////////////////////%

% obs status
The recent observation of gravitational waves (GWs) by the advanced laser interferometer (aLIGO) and advanced VIRGO (aVIRGO) has opened a new window to probe unseen Universe \cite{LIGO:GW150914}. The aLIGO and aVIRGO detected the ten events from binary black hole (BBH) mergers and one event from a binary neutron star (BNS) merger during the first and second observing runs \cite{LIGO:2018:gwtc}. The third observing run has provided dozens of candidates of GW events, most of these are likely to have originated from BBHs. \footnote{https://gracedb.ligo.org/superevents/public/O3/} In coming years, KAGRA \cite{KAGRA:2013}, located in Kamioka, Japan, will join a network of GW observatories, and a large number of GW sources would be detected. The additional GW detector will also significantly improve the sky localization of GW events. 

% cosmology with luminosity distance
From the cosmological point of view, future GW observations will provide the luminosity distance to each binary source with an unprecedented precision. With the observed luminosity distance, we will be able to precisely constrain cosmic expansion history and cosmology \cite{Schutz:1986,Cutler:2009:GWcosmo,Nissanke:2009,Sathyaprakash:2009:ET,Nishizawa:2011:v,Petiteau:2011,Taylor:2012,Arabsalmani:2013:gw,DelPozzo:2015:gw,Tamanini:2016:gw} (see also \cite{Cai:2017:gw} and references therein). Since the luminosity distance is distorted by the lensing of the foreground large-scale structure, the measured luminosity distance can be also used as a probe of the large-scale structure \cite{Cutler:2009:GWcosmo,Hirata:2010:GW,Shang:2010,Camera:2013:GWsiren}. Alternatively, multiple studies have discussed distribution of GW sources as a tracer of the large-scale structure. We can constrain cosmology and progenitors of GW sources by cross-correlating GW sources with galaxy catalogs \cite{Oguri:2016:GW,Namikawa:2016:gwc,Raccanelli:2016:GW,Scelfo:2018:gw,Mukherjee:2019:gw,Calore:2020:gw}, gravitational lensing \cite{Namikawa:2015:gw,Namikawa:2016:gwc,Osato:2018:gw}, or by the correlation function and angular power spectrum of GW sources \cite{Namikawa:2015:gw,Namikawa:2016:gwc,Palmese:2020:GW}. Recent studies by \cite{Cavaglia:2020:gw,Payne:2020:gw} analyze the GW sources obtained from the first and second observing runs of aLIGO/aVIRGO, but a clustering signal of the GW sources has not yet been detected. In the future, however, the clustering signal would be detectable and can be used to constrain Hubble parameter \cite{Oguri:2016:GW,Osato:2018:gw,Zhang:2018:GWPS}, primordial non-Gaussianity \cite{Namikawa:2015:gw}, and progenitors of BBHs and primordial black holes \cite{Raccanelli:2016:GW,Scelfo:2018:gw}. 

% redshift determination
It is, however, not realistic to assume that we will be able to observe redshift for each GW source in analysis of the GW source clustering, because GW observations alone are not sensitive to source redshifts. Redshift identification of BNSs with electromagnetic follow-up observations is rather challenging, in particular, at high redshifts \cite{Nishizawa:2011:GW,Nissanke:2012dj,Nishizawa:2019,Yu:2020:gw-strong-lens}, and several alternative methods to infer the redshift depend largely on the assumptions \cite{MacLeod:2007:GW,Messenger:2011,Taylor:2011,Safarzadeh:2019:gw}. Source redshift identification for BBHs by a galaxy catalogue is limited to low $z$ sources \cite{Kyutoku:2016:hubble,Nishizawa:2016:hubble}. The limitation of available sources for clustering analysis leads to significant loss of signal-to-noise. A simple yet optimal way is, therefore, to use observed luminosity distance as a distance indicator for each GW source \cite{Namikawa:2015:gw}. With this approach, multiple studies have so far explored the feasibility of detecting the clustering signal of GW sources and its application to cosmology (e.g. \cite{Oguri:2016:GW,Namikawa:2016:gwc,Osato:2018:gw,Zhang:2018:GWPS,Mukherjee:2020:gw,Bera:2020jhx}). Instead of employing the resolved GW sources, multiple studies have discussed mapping of stochastic GW backgrounds from cosmological and astrophysical sources as similar to the anisotropies of the cosmic microwave backgrounds  \cite{Allen:1996:gw,Cornish:2001:gw,Mitra:2007:gw,Thrane:2009:gw} (see also recent works by e.g. \cite{Romano:2015:gw,Contaldi:2016:GW,Cusin:2017:GW,Jenkins:2018:gw,Jenkins:2019:GW}). The expected noise in the map obtained from the forthcoming GW detectors is, however, several orders of magnitude larger than the clustering signals \cite{Alonso:2020:GW}. 

In this paper, we extend the studies of \cite{Namikawa:2015:gw,Namikawa:2016:gwc} to the three-dimensional number density of GW sources, its correlation function and multipole power spectrum. The correlation function and power spectrum of the density fluctuations have been widely used for analysis in galaxy surveys and will be useful for analysing the clustering of GW sources. 
Using observed luminosity distance and direction of GW sources, we can estimate source positions in the comoving space and compute density contrast of the GW sources. Source peculiar velocities and foreground lensing distort observed luminosity distance, and the observed number density, correlation function and power spectrum are all distorted as similar to those in galaxy surveys. Recently, \cite{Zhang:2018:GWPS} makes an ansatz that the distortions are induced by source peculiar velocities alone and obtains a similar anisotropic power spectrum to that derived in the redshift-space case. In this paper, however, we find that a line-of-sight derivative of the lensing convergence, which does not appear in the redshift-space distortions and ignored in the previous work, is important to characterize the observed number density fluctuations. We also clarify the difference of the correlation function and power spectrum from galaxy redshift surveys. \cite{Vijaykumar:2020:gw,Cavaglia:2020:gw,Payne:2020:gw,Mukherjee:2020:gw,Bera:2020jhx} discuss the correlation function and power spectrum of the GW sources but the distortions derived in this paper are not explored. The distortion effect, however, must be corrected even if we use a redshift transformed from the observed luminosity distance by assuming cosmology for future GW observations. 

This paper is organized as follows. In sec.~\ref{sec:lds}, we derives the number density fluctuations obtained from the observed luminosity distance and sky location of GW sources. Sec.~\ref{sec:pk} shows the correlation function and multipole power spectrum of the observed density fluctuations, and the cross correlation between GW sources and galaxies. Sec.~\ref{sec:summary} is devoted for summary and discussion. 

Throughout this paper, we assume the flat $\Lambda$CDM cosmology and fiducial parameters are consistent with the latest Planck results \cite{P18:main}; the fractional energy density of baryon $\Omega_b=4.4\times10^{-2}$, of matter $\Omega_{\rm m}=0.30$, Hubble constant $H_0=70$~km/s/Mpc, primordial scalar spectral index $n_{\rm s} = 0.96$, and primordial scalar amplitude $A_{\rm s}=2.2\times10^{-9}$ at $k_0=0.05$Mpc$^{-1}$. We use the halofit non-linear matter power spectrum obtained from {\tt CAMB} \cite{Lewis:1999bs}.

%//////////////////////////////////////////////////////////////////////////////////////////%
\section{Density Fluctuations in Luminosity-Distance Space} \label{sec:lds}
%//////////////////////////////////////////////////////////////////////////////////////////%

\subsection{Luminosity-Distance Space Distortions}

In observations, we measure the luminosity distance and arrival direction of a GW. For a given cosmology, both the comoving distance and luminosity distance are explicitly given as a function of the redshift, and the comoving distance is considered as a function of the luminosity distance. Therefore, we can convert the luminosity distance and arrival direction to the comoving coordinate position, $\bs$. We call this coordinate system, $\bs$, ``luminosity-distance space'', analogues to the ``redshift space'' in galaxy surveys. 

The luminosity distance is perturbed by the peculiar velocity of sources and lensing \cite{Sasaki:1987}. The comoving distance derived from the observed luminosity distance is then modified and the spatial distributions of sources are distorted in the luminosity distance space. To derive the distortion effect on the observed source density fluctuations, we consider observed luminosity distance given by $D+\delta D$ where $D$ is the luminosity distance in the unperturbed universe and $\delta D\equiv D\epsilon$ is a small perturbation. The comoving distance computed from the observed luminosity distance is then modified as: 
%------------------------------------------------------------------------------------------%
\al{
    s = r(D+\delta D) \simeq r(D) + \D{r}{D}\delta D \equiv r+r\wte 
    \,, \label{Eq:s-r}
}
%------------------------------------------------------------------------------------------%
where $r$ is the comoving distance as a function of $D$. Using the conformal Hubble, $\mC{H}$, and $r_H=r\mC{H}$, we also define: 
%------------------------------------------------------------------------------------------%
\al{
    \wte \equiv \frac{\epsilon}{1+r_H}
    \,. \label{Eq:ds}
}
%------------------------------------------------------------------------------------------%
The perturbations to the luminosity distance from the large-scale structure are given by (e.g. \cite{Sasaki:1987,Hui:2005:LD}): 
%------------------------------------------------------------------------------------------%
\al{
    \epsilon = 2v_r - \kappa 
    \,, \label{Eq:delta-d}
}
%------------------------------------------------------------------------------------------%
where we ignore the gravitational potential perturbations (except lensing) and monopole and dipole components induced by the perturbations at the observer position. In observations, the monopole and dipole terms only introduce anisotropies at the largest scales, $\l\leq 1$. We are interested in the sub-horizon scale density fluctuations and ignore these terms in the followings. 

Note that a small difference in the luminosity distance, $\Delta D$, and angular separation, $\Delta\theta$, corresponds to a small comoving displacement along and perpendicular to a line-of-sight as follows: 
%------------------------------------------------------------------------------------------%
\al{
    \Delta s_{\|} &= \frac{a}{1+r_H} \Delta D
    \,, \\
    \Delta s_{\perp} &= r\Delta\theta
    \,, 
}
%------------------------------------------------------------------------------------------%
where $a$ is the scale factor. The above conversion depends on cosmology and the relationship can be used to test assumed cosmology. This is analogues to the so-called Alcock-Paczynski test in the redshift galaxy survey \cite{Alcock:1979:APtest}. 

Using the number conservation law, the density fluctuations in the luminosity-distance space are related to that in the comoving space as: 
%------------------------------------------------------------------------------------------%
\al{
    1+\delta_s = \frac{\ncsrc(r)}{\ncsrc(s)}\bigg|\D{^3s}{^3r}\bigg|^{-1}
    (1+\delta_r)
    \,. 
}
%------------------------------------------------------------------------------------------%
Substituting \eq{Eq:s-r} into the above equation and expanding the equation up to linear order of perturbations, we find: 
%------------------------------------------------------------------------------------------%
\al{
    1+\delta_s 
    &= \frac{\ncsrc(r)}{\ncsrc(r+r\wte)}
    \frac{r^2}{(r+r\wte)^2}\left(\D{}{r}[r(1+\wte)]\right)^{-1}
    (1+\delta_r)
    \\ 
    &\simeq \left(1-3\wte-\wte\D{\ln\ncsrc(r)}{\ln r}-r\D{\wte}{r}\right) 
    (1+\delta_r)
    \,. 
}
%------------------------------------------------------------------------------------------%
In addition, the lensing distortion to the angular position additionally leads to $-2\kappa$ in the above equation as similar to redshift surveys (see Appendix \ref{app:full} for a derivation). Substituting \eq{Eq:ds} into the above equation and adding $-2\kappa$, we find that the density fluctuations in the luminosity distance space are distorted as: 
%------------------------------------------------------------------------------------------%
\al{
    \delta_s = \delta_r - \left(\alpha + \frac{\gamma}{\mC{H}}\D{}{r}\right) \epsilon 
    -2\kappa
    \,. \label{Eq:delta-s}
}
%------------------------------------------------------------------------------------------%
where we define: 
%------------------------------------------------------------------------------------------%
\al{
    \gamma &\equiv \frac{r_H}{1+r_H} 
    \,, \\
    \alpha &\equiv \frac{1}{1+r_H} 
    \left(3+\D{\ln n_{\rm s}(r)}{\ln r}-\gamma+\frac{\gamma r_H\mC{H}_{,\eta}}{\mC{H}^2}\right)
    \,. \label{Eq:alpha}
}
%------------------------------------------------------------------------------------------%
Here, $\mC{H}_{,\eta}\equiv{\rm d}\mC{H}/{\rm d}\eta=-{\rm d}\mC{H}/{\rm d}r$ is the derivative of the conformal Hubble with respect to the conformal time, $\eta$. The perturbations, $\epsilon$, contain the radial peculiar velocity and lensing convergence. Substituting \eq{Eq:delta-d} into \eq{Eq:delta-s}, we find:
%------------------------------------------------------------------------------------------%
\al{
    \delta_s = \delta_r - 2\alpha v_r - \frac{2\gamma}{\mC{H}}\D{v_r}{r}  
    + \left(-2+\alpha+\frac{\gamma}{\mC{H}}\D{}{r}\right)\kappa
    \,. \label{Eq:delta-s-lead}
}
%------------------------------------------------------------------------------------------%

%------------------------------------------------------------------------------------------%
\begin{figure}
\centering
\includegraphics[width=80mm]{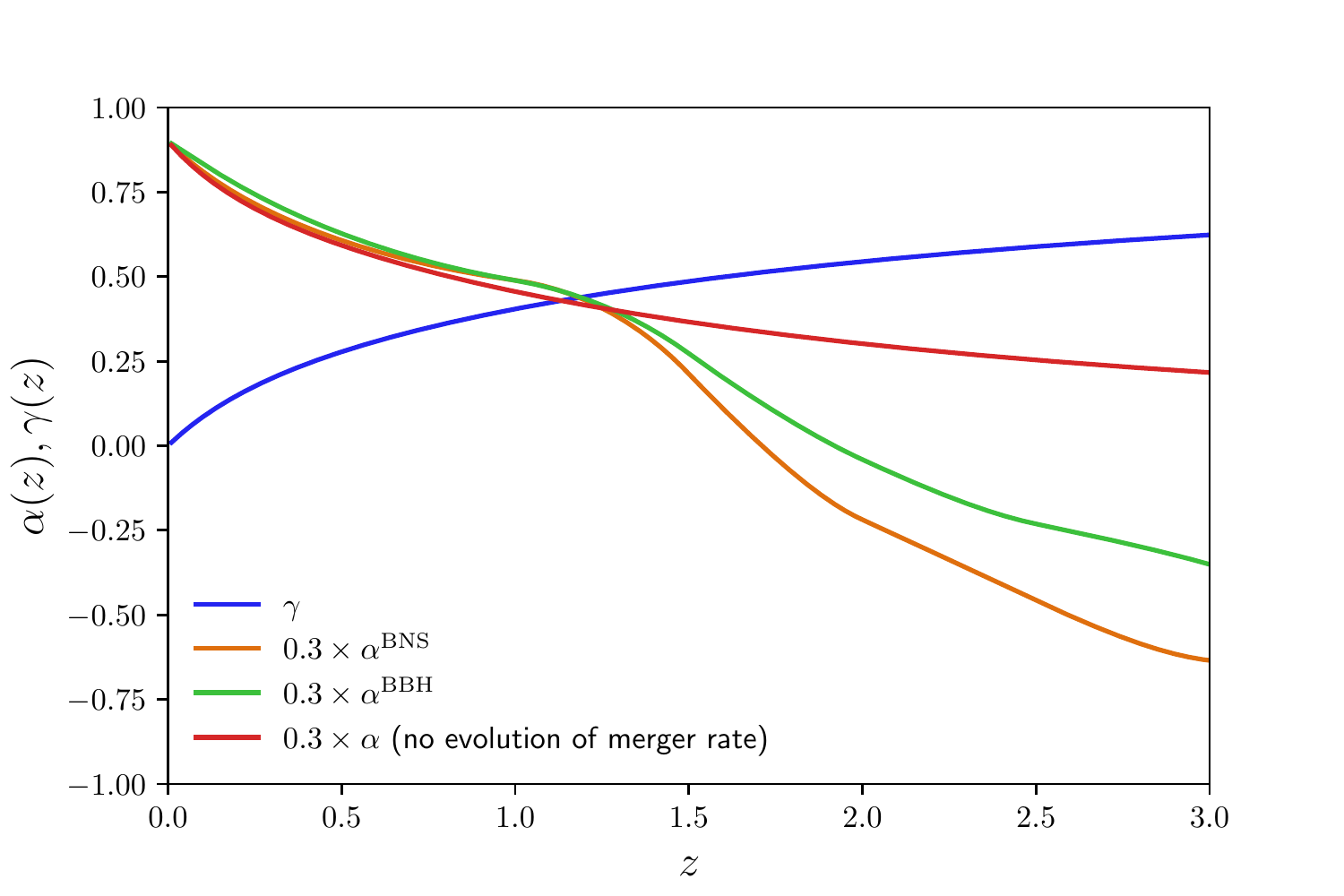}
\caption{
Two new coefficients for the luminosity-distance space distortions, $\gamma$ and $\alpha$, as a function of $z$. For $\alpha$, we also show the case when the merger rate does not evolve. For illustration, $\alpha$ is multiplied by $0.3$.
}
\label{fig:coeff}
\end{figure}
%------------------------------------------------------------------------------------------%

The luminosity distance perturbations from the source radial velocity is related to the matter density fluctuations. Using the Fourier modes of the matter density fluctuations, $\delta_{\rm m}(\eta,\bk)$, and assuming the linear constant bias, $b$, the density and velocity (non-lensing) contributions are given by: 
%------------------------------------------------------------------------------------------%
\al{
    \delta_s|^{\delta+v}(\bs) = \Int{3}{\bk}{(2\pi)^3}\E^{\iu\bk\cdot\bs}
    \left(1+ 2\gamma \frac{f}{b} \mu^2 \right) b\delta_{\rm m}(\eta(s),\bk)
    \,, \label{Eq:delta:d+v}
}
%------------------------------------------------------------------------------------------%
where $\mu$ is the cosine between $\bs$ and $\bk$, and $f$ is the linear growth rate. Note that we apply the sub-horizon limit and ignore the terms at $\mC{O}(1/k_H)$ where $k_H=k/\mC{H}$, but the distant observer approximation is not applied. We ignore higher-order perturbations. Thus, in the argument of the density fluctuations, the true comoving space position, $\br$, equals to the luminosity-distance space position, $\bs$. The time-dependent quantities are also evaluated by the observed comoving distance, $s$. We approximate ${\rm d}/{\rm d}r\simeq \pd/\pd r$ to focus on the sub-horizon scale. The above density fluctuations have a similar form to that derived in the redshift space \cite{Kaiser:1987}. At low redshifts, the factor $\gamma$ goes to zero and the quadrupole term becomes negligible. 

For a given comoving distance to a source, $r$, and direction, $\hatn$, the lensing convergence and its derivative are given as (see e.g. \cite{Bartelmann:2001,Munshi:2008:review}): 
%------------------------------------------------------------------------------------------%
\al{
    \kappa(r,\hatn) &= \INT{}{r'}{}{0}{r}\frac{(r-r')r'}{r} \bn^2_{\hatn}\psi(r',r'\hatn)
    \,, \label{Eq:kappa} \\
    \D{\kappa(r,\hatn)}{r} &= \INT{}{r'}{}{0}{r}\frac{r'^2}{r^2} \bn^2_{\hatn}\psi(r',r'\hatn)
    \,, 
}
%------------------------------------------------------------------------------------------%
where $\bn_{\hatn}$ is the angular components of the covariant derivative on a sphere of radius $r'$ and $\psi$ is the Weyl potential. The total lensing contribution to the density fluctuations is obtained by substituting the above equations into \eq{Eq:delta-s} with $\epsilon=-\kappa$, yielding: 
%------------------------------------------------------------------------------------------%
\al{
    \delta_s|^{\kappa}(\bs) = \INT{}{r'}{}{0}{\infty} 
    w(s,r') \bn^2_{\hatn}\psi(\eta(r'),r'\hatn) 
    \equiv \mC{K}(s,\hatn) 
    \,, \label{Eq:lens-term}
}
%------------------------------------------------------------------------------------------%
where we introduce a modified lensing kernel: 
%------------------------------------------------------------------------------------------%
\al{
    w(r,r') = \frac{r'}{r}\left[(-2+\alpha)(r-r')+\frac{\gamma}{\mC{H}}\frac{r'}{r}\right] 
    \,, 
}
%------------------------------------------------------------------------------------------%
if $0\leq r'\leq r$, and $0$ otherwise. 

Fig.~\ref{fig:coeff} shows the redshift dependence of two coefficients, $\gamma$ and $\alpha$. To evaluate $\alpha$, we need the number density of the GW sources per comoving volume \cite{Cutler:2005:BBO,Nishizawa:2011:GW,Namikawa:2016:gwc}: 
%------------------------------------------------------------------------------------------%
\al{
    n_{\rm s}(r) &= \frac{T_{\rm obs}\dot{n}_0\mC{R}[z(r)]}{1+z(r)} 
    \,, \label{Eq:ns-z}
}
%------------------------------------------------------------------------------------------%
where $\dot{n}_0$ is the merger rate today and $\mC{R}(z)$ encapsulates the time evolution of the rate. We compute the time dependence of the coefficient, $\alpha$, by substituting \eq{Eq:ns-z} into \eq{Eq:alpha}. We adopt the time evolution of the rate $\mC{R}(z)$ derived in \cite{Safarzadeh:2019:gw} (with $t_{\rm min}=100$\,Myr, $\Gamma=-1$) for BNSs and \cite{Chen:2019:gw} for BBHs, respectively. The factor $2\gamma$ becomes larger than unity at $z\agt 1.3$. $\alpha$ becomes large at high redshifts due to the time dependence of the merger rate. 

Finally, the density fluctuations in the luminosity-distance space are given by combining \eqs{Eq:delta:d+v,Eq:lens-term}, yielding: 
%------------------------------------------------------------------------------------------%
\al{
    \delta_s(\bs) = \Int{3}{\bk}{(2\pi)^3}\E^{\iu\bk\cdot\bs}
    \left(1+2\gamma\beta\mu^2\right) b\delta_{\rm m}(\eta(s),\bk)
    + \mC{K}(s,\hatn)
    + \mC{O}(1/k_H) 
    \,, \label{Eq:delta-all-kH}
}
%------------------------------------------------------------------------------------------%
where $\beta=f/b$ and $\mC{K}$ is defined in \eq{Eq:lens-term}. The quadrupole term arising from the peculiar velocity is consistent with \cite{Zhang:2018:GWPS}. In Appendix \ref{app:full}, we derive the distortions in a more rigorous approach based on linear theory of General Relativity, finding that \eq{Eq:delta-all-kH} contains all of the terms important in the sub-horizon scales. 

\subsection{Comparison with Redshift Space Distortions}

The above derivation is analogues to that used for the redshift-space distortion. In galaxy surveys, positions of each galaxy in comoving coordinate are derived from the observed redshift and direction (see e.g. \cite{Hamilton:RSD:review,Kaiser:1987}). The observed redshift-space distance $s$ differs from the distance in real space $r$ by the comoving radial velocity displacement:
%------------------------------------------------------------------------------------------%
\al{
    s = r + \frac{v_r}{\mC{H}} 
    \,. 
}
%------------------------------------------------------------------------------------------%
Using the galaxy number conservation, the density fluctuations in the redshift space are related to that in the real space as (see e.g. \cite{Matsubara:2000:rsd}): 
%------------------------------------------------------------------------------------------%
\al{
    \delta^{\rm gal}_s \simeq \delta^{\rm gal}_r 
    - \left(\frac{\alpha'}{s}+\D{}{s}\right) \frac{v_r}{\mC{H}} 
    + (5s_{\rm g}-2)\kappa
    \,.
}
%------------------------------------------------------------------------------------------%
where $\alpha'=2+{\rm d}\ln n_{\rm g}/{\rm d}\ln s$ and $s_{\rm g}$ is the logarithmic slope of the number counts. Assuming the linear constant galaxy bias, $b'$, the density fluctuations in the redshift space are recast as: 
%------------------------------------------------------------------------------------------%
\al{
    \delta^{\rm gal}_s(\bs) = \Int{3}{\bk}{(2\pi)^3} \E^{\iu\bk\cdot\bs} 
    \left(b'-\iu\mu\frac{\alpha' f}{ks}+f\mu^2\right)\delta_{\rm m}(\eta(s),\bk)
    + (5s_{\rm g}-2)\kappa(s,\hatn)
    \,. \label{Eq:delta-all-kH-gal}
}
%------------------------------------------------------------------------------------------%
In the full general relativistic description, there are several terms at $\mC{O}(1/k_H)$ (see e.g. Eq.~(4.18) of \cite{Yoo:2014:GR}). One notable difference is that \eq{Eq:delta-s} contains the distortion from the line-of-sight derivative of the lensing convergence. As we will see below, the line-of-sight derivative of the lensing convergence significantly modifies the correlation function when the separation vector is almost parallel to the line-of-sight direction. Another difference between the redshift space and luminosity-distance space distortions is the dipole term arising from the radial peculiar velocity of sources. In the context of the redshift-space distortion, the dipole term is negligible if $ks\gg 1$. In the luminosity-distance space distortion, on the other hand, the dipole term vanishes in the sub-horizon limit, $k_H\gg 1$. 

%::::::::::::::::::::::::::::::::::::::::::::::::::::::::::::::::::::::::::::::::::::::::::%
\section{Correlation Function and Multipole Power Spectrum} \label{sec:pk}
%::::::::::::::::::::::::::::::::::::::::::::::::::::::::::::::::::::::::::::::::::::::::::%

To extract cosmological and astrophysical information from the clustering of the GW sources, we measure statistics such as the two-point correlation function and power spectrum of the density fluctuations. Here, we explore how the correlation function and multipole power spectrum are distorted in the luminosity-distance space. In the followings, we assume $b(z)=\sqrt{1+z}$ \cite{Namikawa:2016:gwc}. 

\subsection{Correlation Function}

We first derive the two-point correlation function of the density fluctuations at positions, $\bs_1$ and $\bs_2$, in the luminosity-distance space. The two-point correlation function is defined as $\xi(\bs_1,\bs_2)\equiv\ave{\delta_s(\bs_1)\delta_s(\bs_2)}$. In the following, we define the separation vector, $\bx=\bs_2-\bs_1$, and the line-of-sight vector, $\bd=\bs_1$, and focus on the sub-horizon limit, $k_H\gg 1$. We adopt the plane-parallel approximation in which the separation is much smaller than the line-of-sight distance, $x\ll d$, and the three directions, $\hs_1$, $\hs_2$, and $\hd$, are almost identical. The correlation function of the density fluctuations at two different positions is then expressed in terms of the separation and line-of-sight vectors as: 
%------------------------------------------------------------------------------------------%
\al{
    \xi(\bx,d) = \Int{3}{k}{(2\pi)^3} 
    \E^{\iu\bk\cdot\bx} (1+2\mu^2\beta\gamma)^2 
    b^2P_{\rm m}(\eta(d),k) 
    + \xi^{\delta\mC{K}}(\bx,d) 
    + \xi^{\mC{K}\mC{K}}(\bx,d)
    \,, \label{Eq:xi-P}
}
%------------------------------------------------------------------------------------------%
where $\mu=\bd\cdot\bk/dk$ and $P_{\rm m}(\eta,k)$ is the matter power spectrum: 
%------------------------------------------------------------------------------------------%
\al{
    P_{\rm m}(\eta,k)(2\pi)^3\delta^D(\bk-\bk') 
    \equiv \ave{\delta_{\rm m}(\eta,\bk)\delta^*_{\rm m}(\eta,\bk')} 
    \,, 
}
%------------------------------------------------------------------------------------------%
with $\delta^D$ being the Dirac delta function in three dimensions. The correlation function between the density and lensing effect and the lensing correlation function in the Limber approximation are obtained by \cite{Matsubara:2000:rsd,Hui:2007:RSDmag-I} but replacing the lensing kernel to $w(r,r')$:
%------------------------------------------------------------------------------------------%
\al{
    \xi^{\delta\mC{K}} (\bx,d) 
    &= \frac{\gamma}{\mC{H}}[\Theta(x_{\|})+\Theta(-x_{\|})]
    \frac{3\Omega_{\rm m}H_0^2}{2a(d)} 
    \Int{}{k_{\perp}}{2\pi} k_\perp J_0(k_\perp x_\perp) bP_{\rm m}(\eta(d),k_{\perp})
    \,, \label{Eq:xi-deltak} \\ 
    \xi^{\mC{K}\mC{K}} (\bx,d) 
    &= \INT{}{r'}{}{0}{\infty} w^2(d,r') 
    \frac{9\Omega^2_{\rm m}H_0^4}{4a^2(r')} 
    \Int{}{k_{\perp}}{2\pi} k_\perp J_0(k_\perp r'x_\perp/d) 
    P_{\rm m}(\eta(r'),k_{\perp})
    \,. \label{Eq:xi-kk}
}
%------------------------------------------------------------------------------------------%
Here, the step function $\Theta(x)$ is unity when $x\geq 0$ and $0$ otherwise. $x_{\|}$ and $x_\perp$ are the length of the separation vector parallel and perpendicular to the line-of-sight, and $k_\perp$ is the length of the wavevector being orthogonal to the line-of-sight vector, $k^2=k^2_{\|}+k^2_\perp$. $J_0$ is the Bessel function of the first kind of order zero. In deriving the density-lensing and lensing-lensing correlation functions, we use $d\gg x_{\|},x_{\perp}$. The density-lensing correlation function is dominated by the term from the line-of-sight derivative of the lensing convergence. In the lensing correlation function, only Fourier modes perpendicular to the line-of-sight contribute to the lensing as similar to the magnification effect in the redshift-space correlation function \cite{Matsubara:2000mw,Hui:2007:RSDmag-I,Tansella:2017:RSDfull}. Therefore, the lensing and velocity correlation vanishes in the Limber approximation. 

%------------------------------------------------------------------------------------------%
\begin{figure}
\centering
\includegraphics[width=75mm]{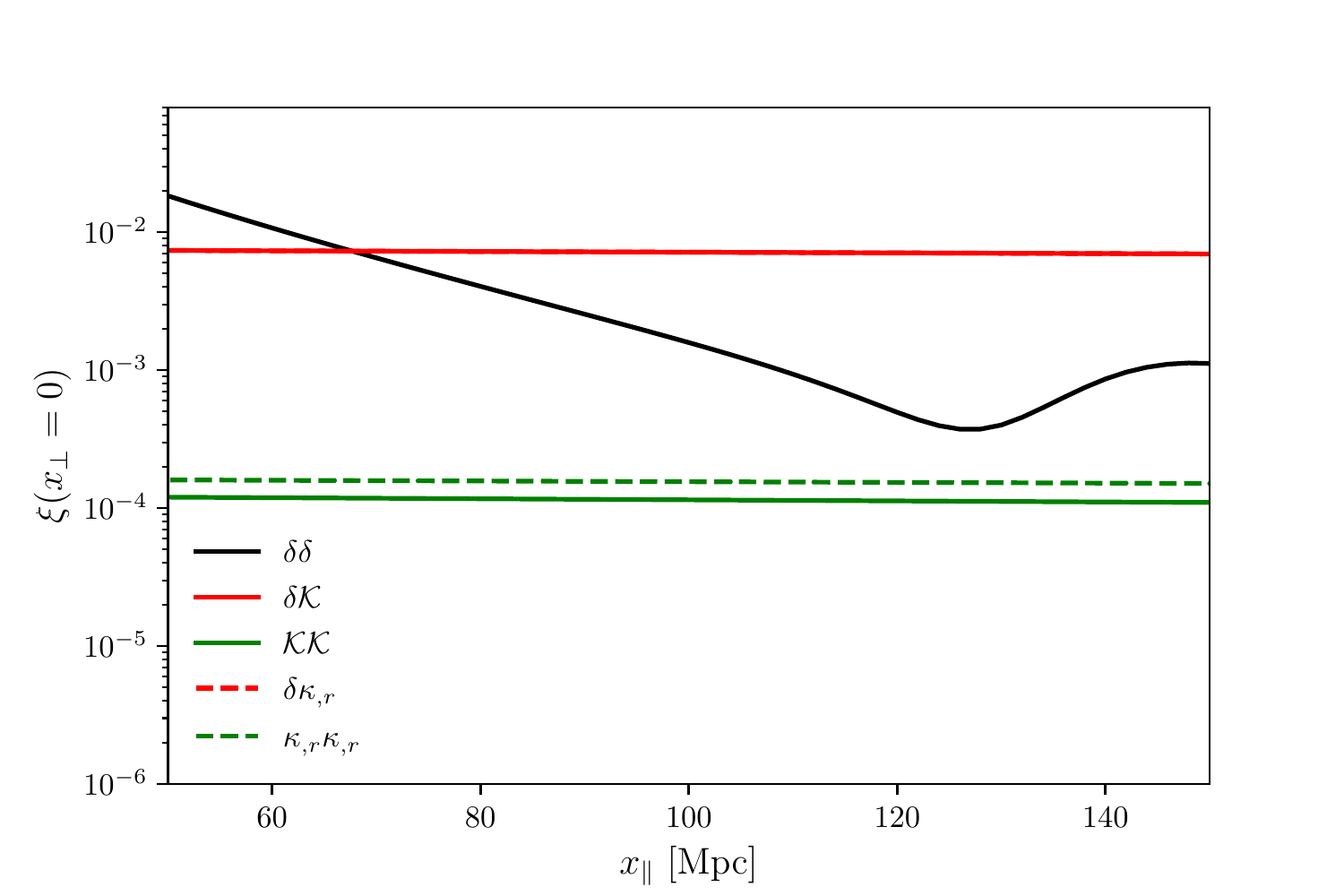}
\includegraphics[width=75mm]{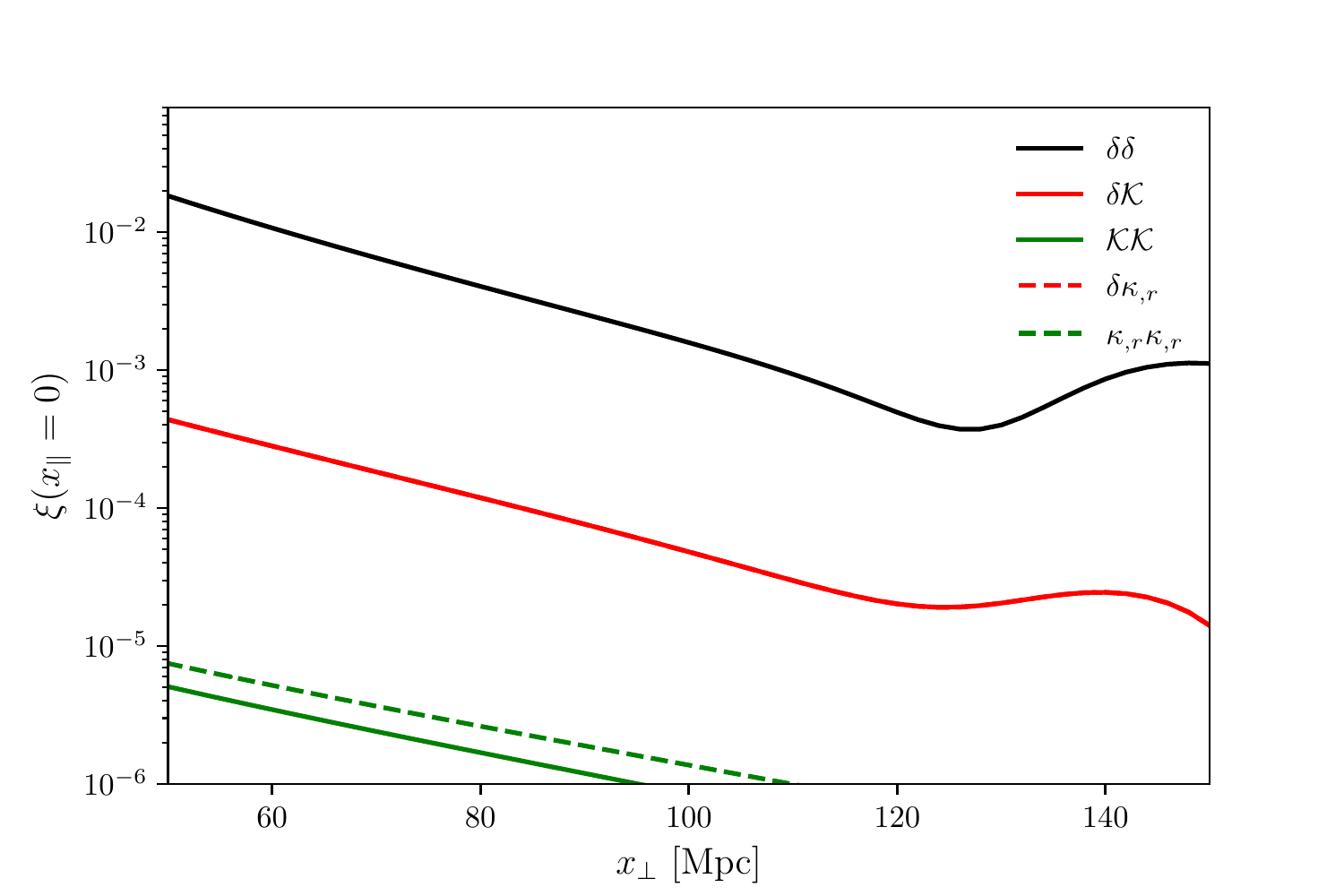}
\includegraphics[width=75mm]{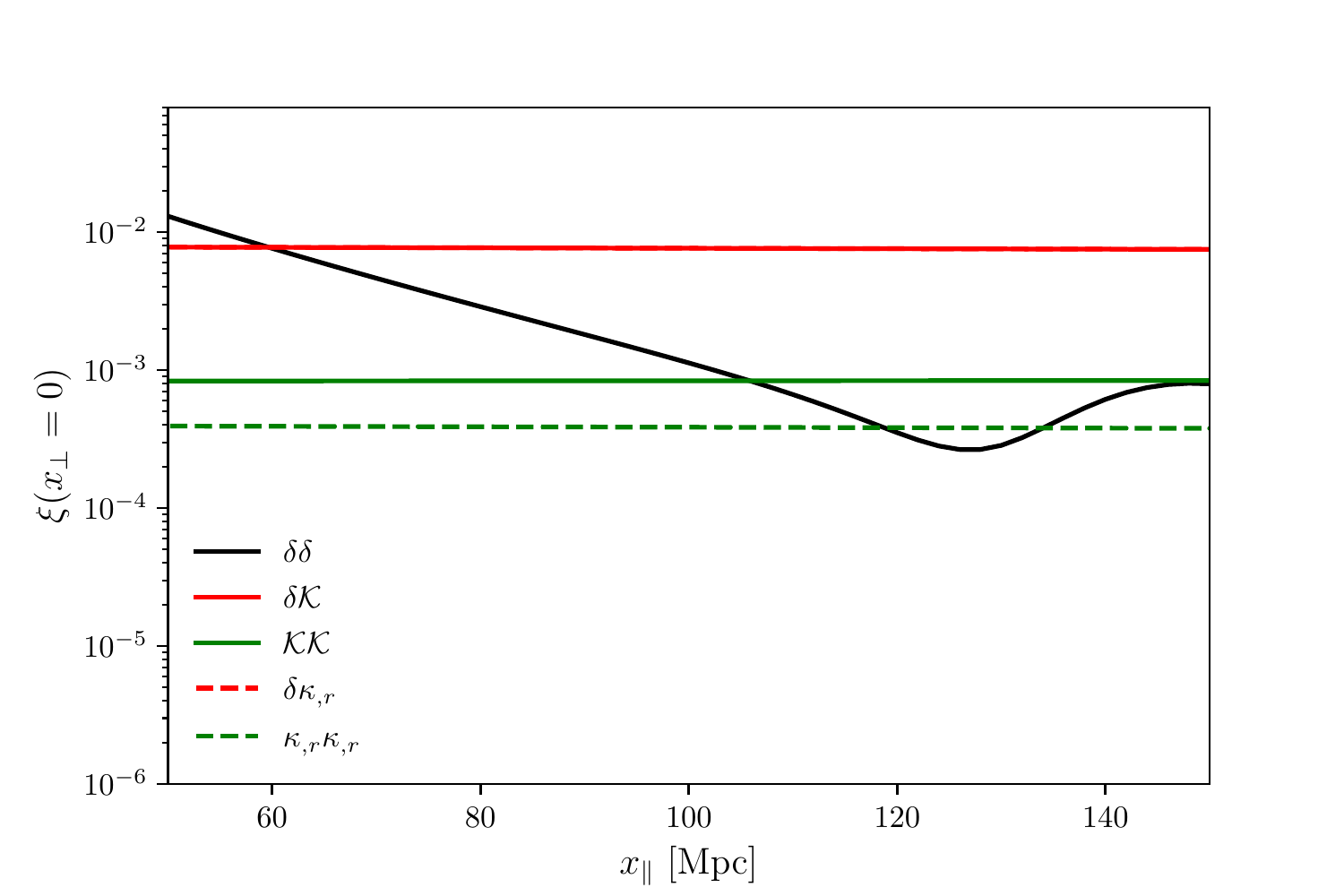}
\includegraphics[width=75mm]{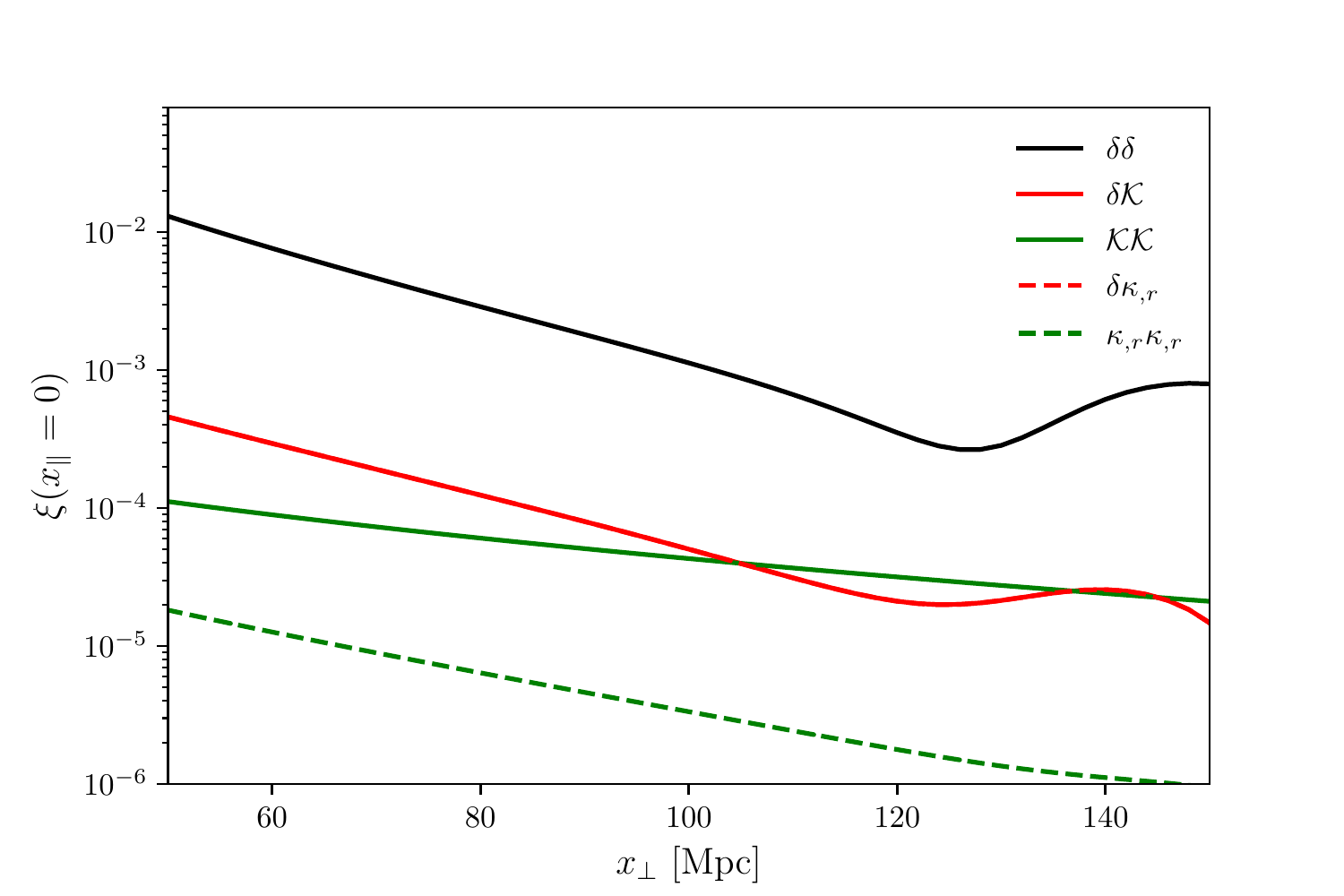}
\caption{
The two-point correlation function of the density fluctuations of BBHs from the density-density ($\delta\delta$), density-lensing ($\delta\mC{K}$) and lensing-lensing ($\mC{K}\mC{K}$) correlations. Top and bottom panels show the correlation function at $D=6.6$\,Gpc (equivalent to $z=1$) and $D=15.5$\,Gpc (equivalent to $z=2$), respectively. We show the two-point correlation function as a function of the line-of-sight separation in left ($x=x_{\|}$ and $x_{\perp}=0$) and of the transverse distance in right ($x=x_\perp$ and $x_{\|}=0$). Dashed lines show the contributions only from the term involving ${\rm d}\kappa/{\rm d}r$.
}
\label{fig:xi_x}
\end{figure}
%------------------------------------------------------------------------------------------%

%------------------------------------------------------------------------------------------%
\begin{figure}
\centering
\includegraphics[width=120mm]{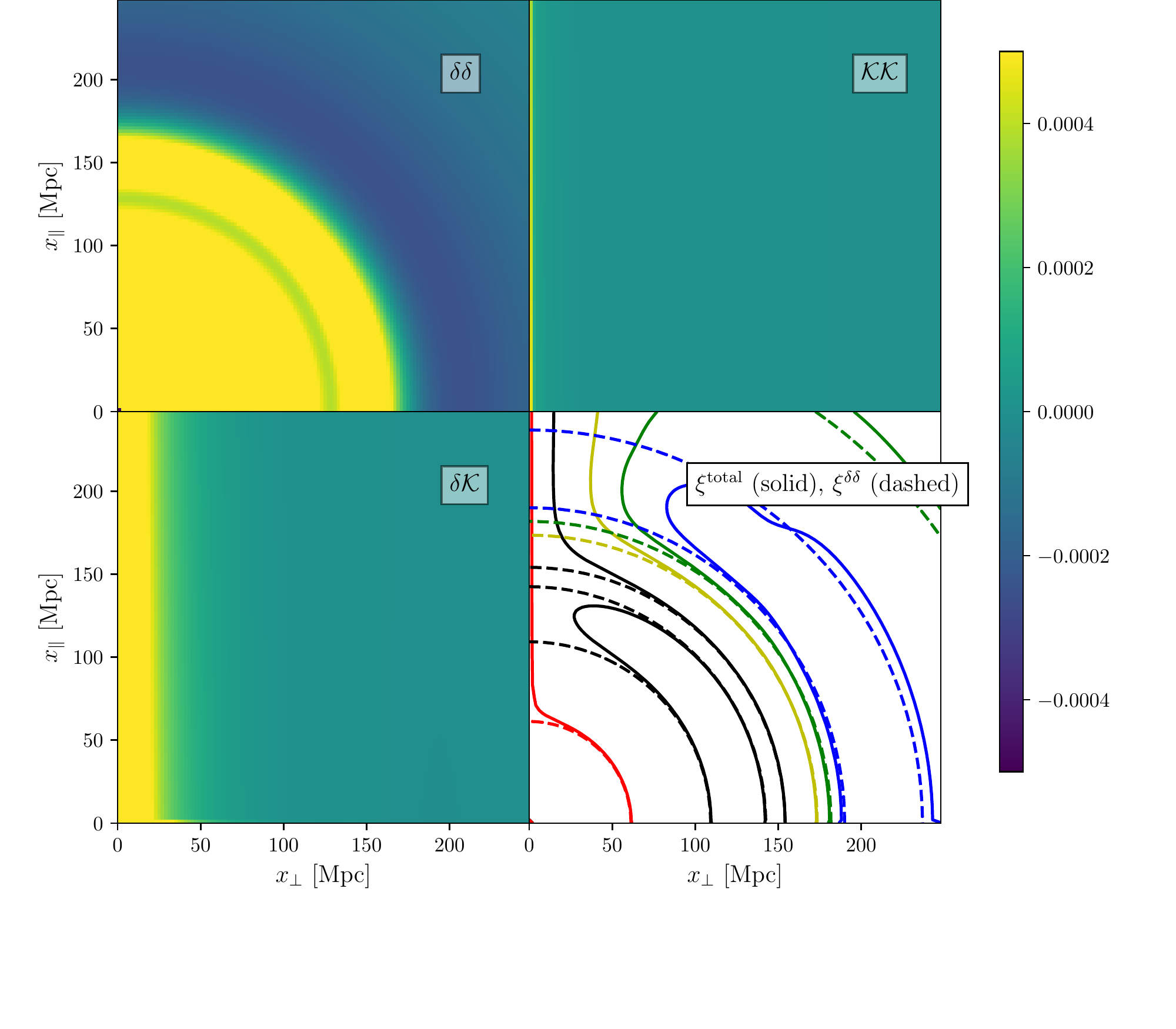}
\caption{
The two-point correlation function of BBHs as a function of $x_{\|}$ and $x_{\perp}$ at $D=6.6$\,Gpc (equivalent to $z=1$) from the density-density (top left), lensing-lensing, (top right) and density-lensing (bottom left) correlations. In the bottom right, we compare the total correlation function (solid contour) with the density-only correlation function (dashed contour). 
} 
\label{fig:xi_zlow}
\end{figure}
%------------------------------------------------------------------------------------------%

%------------------------------------------------------------------------------------------%
\begin{figure}
\centering
\includegraphics[width=120mm]{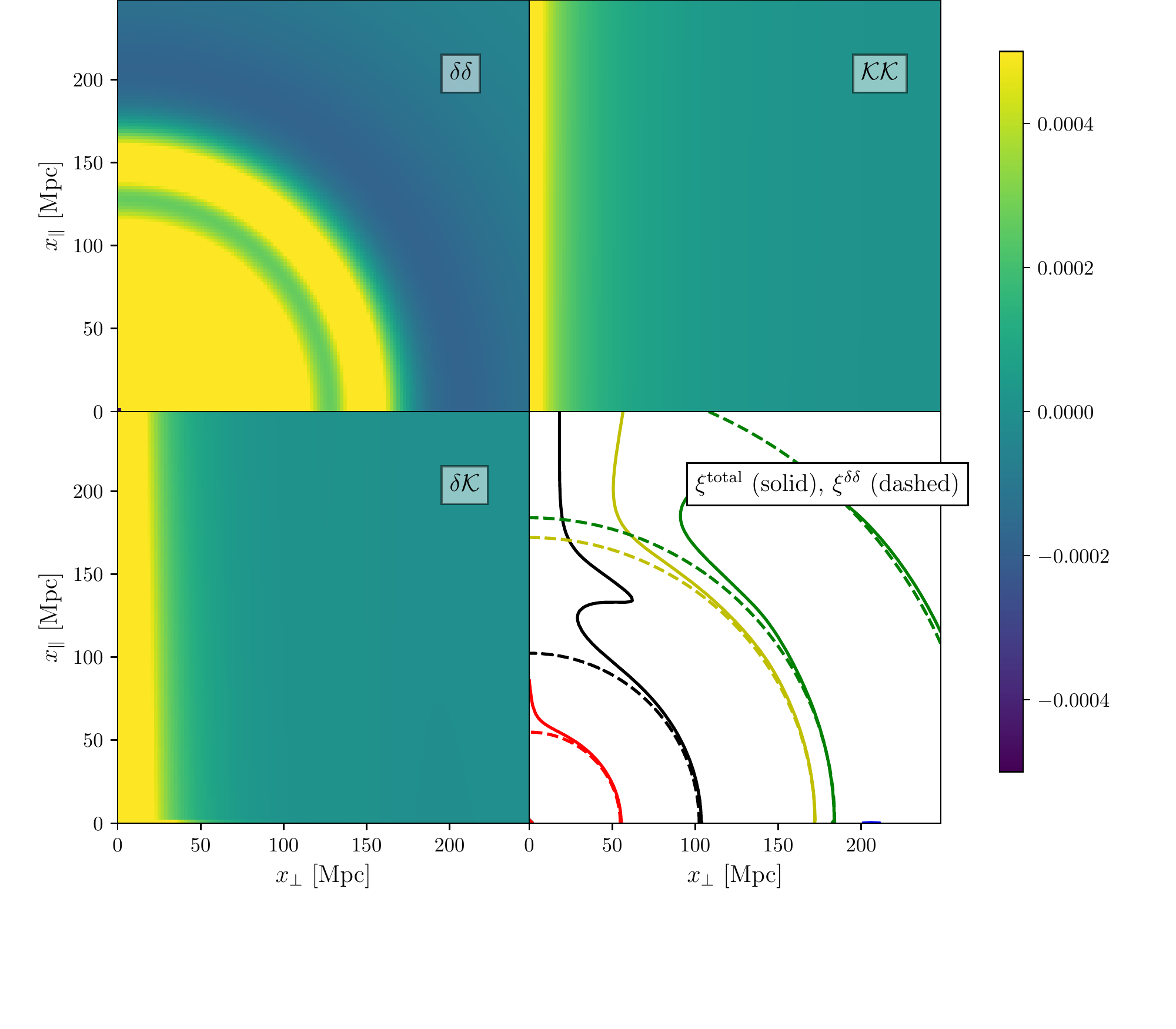}
\caption{
Same as Fig.~\ref{fig:xi_zlow} but for the correlation function at $D=15.5$\,Gpc (equivalent to $z=2$).
}
\label{fig:xi_zmid}
\end{figure}
%------------------------------------------------------------------------------------------%

Fig.~\ref{fig:xi_x} shows the correlation function of BBHs as a function of $x_{\perp}$ ($x_{\|}=0$) or $x_{\|}$ ($x_{\perp}=0$). We do not include the velocity contributions since its impact on the correlation function is identical to that in the redshift space except the additional factor $2\gamma$ which suppresses the quadrupole anisotropies at lower redshifts. We isolate the contribution of the derivative of the lensing convergence which is shown in dashed lines. Note that the redshift dependence of the quantities are ignored when varying $x_{\perp}$ and $x_{\|}$, and all time-dependent quantities are evaluated at $d=r_1$, as similar to \cite{Hui:2007:RSDmag-I}. If the separation vector is approximately parallel to the line-of-sight direction, the dominant contribution comes from the correlation between the density fluctuations and derivative of the lensing convergence. On the other hand, if the separation vector is orthogonal to the line-of-sight direction, the lensing contributions are an order of magnitude smaller than the density auto correlation. 

Figs.~\ref{fig:xi_zlow} and \ref{fig:xi_zmid} show the contour of the correlation function as a function of the separation vector at $z=1.0$ and $2.0$, respectively. We find that, at lower redshift ($z\alt 1$), the correlation between the derivative of the lensing convergence and density fluctuations ($\delta\mC{K}$) introduces a significant anisotropic distortion in the total correlation function. On the other hand, at higher redshift ($z\agt 2$), the lensing auto-correlation ($\mC{K}\mC{K}$) becomes also important.

\subsection{Errors in the Luminosity Distance Measurement and Localization} \label{sec:errors}

%------------------------------------------------------------------------------------------%
\begin{figure}[t]
\centering
\includegraphics[width=80mm]{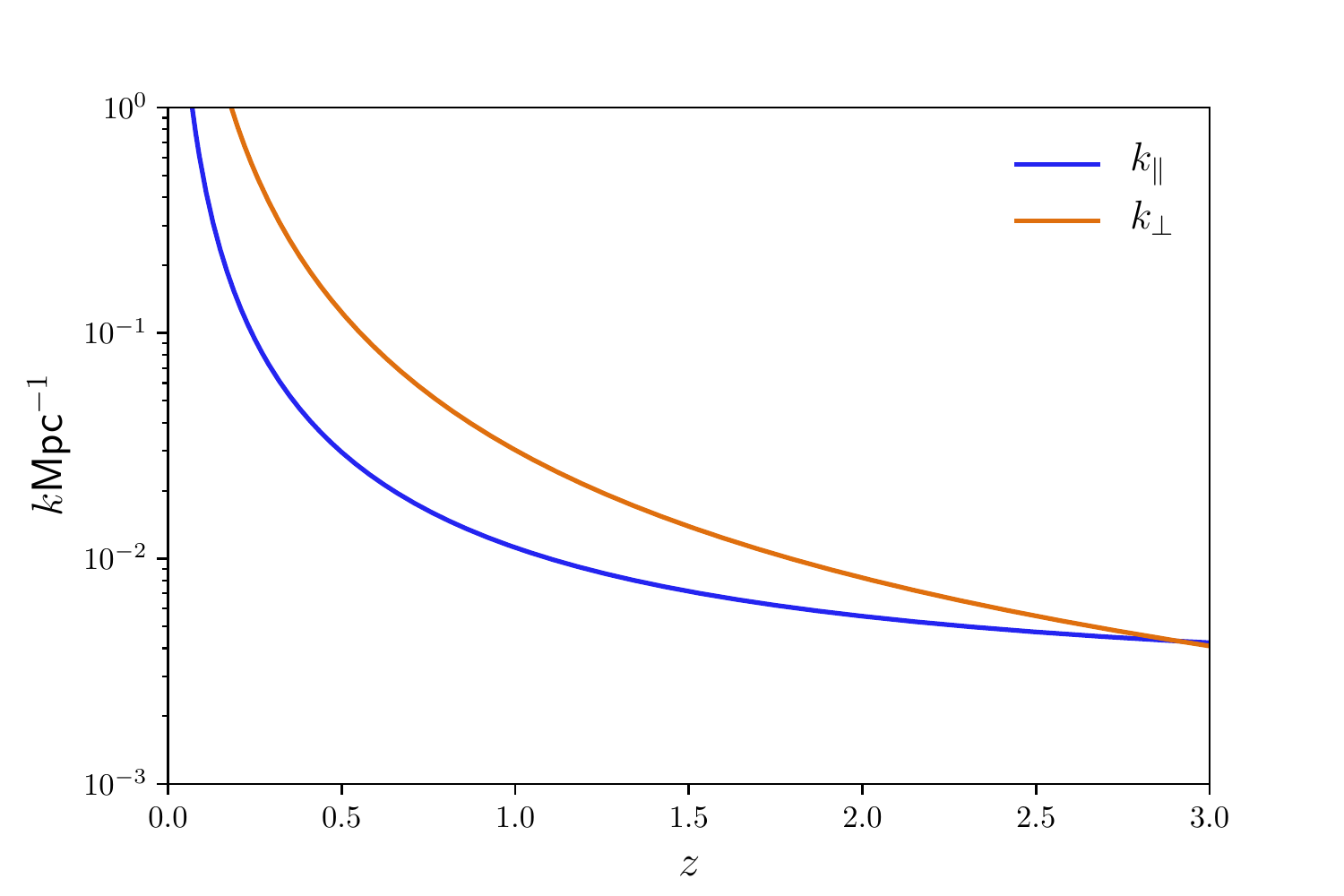}
\caption{
Wavenumers defined by $k_{\|}=(1+d\mC{H})/d\sigma_d$ and $k_{\perp}=1/d\sigma_\theta$ as a function of $z$.The errors are obtained in \cite{Camera:2013:GWsiren,Sidery:2014,Namikawa:2015:gw} by assuming a third generation GW detector network for BNSs.
}
\label{fig:kmax}
\end{figure}
%------------------------------------------------------------------------------------------%

Before moving to the multipole power spectrum, we discuss the impact of measurement errors of the luminosity distance and sky position on the Fourier modes detectable from a observation (see Appendix \ref{app:error} for derivations). 

Supposing that the luminosity distance and sky position have errors $D\to D(1+\epsilon_d)$ and $\hs\to\hs+\delta\hs$, respectively, the observed density fluctuations at scales much smaller than typical size of errors are washed out by the randomness of the errors. In the case of the density and velocity correlation function, from \eq{Eq:xi-P}, these errors introduce an exponential factor, $\ave{\E^{\iu\bk\cdot\delta\bx}}$, in the integrand of $\bk$ where $\delta\bx$ is an error in the separation vector and the average is taken over GW sources. We assume that the errors are described by random Gaussian fields whose standard deviations are $\sigma_d$ for the luminosity distance and $\sigma_\theta$ for angular position, yielding: 
%------------------------------------------------------------------------------------------%
\al{
    \xi^{\delta+v}(\bx,d) 
    &= \Int{3}{\bk}{(2\pi)^3} \E^{\iu\bk\cdot\bx}
    \exp\left[-\frac{(k_{\|}d)^2\sigma_d^2}{(1+d\mC{H})^2}\right]
    \exp\left[-(k_\perp d)^2\sigma_\theta^2\right]
    (1+2\mu^2\beta\gamma)^2 b^2 P_{\rm m}(\eta(d),k)
    \,. \label{Eq:error:d-v}
}
%------------------------------------------------------------------------------------------%
The density-lensing and lensing-lensing correlations do not have contribution from $k_{\|}$ but are suppressed by the localization error. From \eqs{Eq:xi-deltak,Eq:xi-kk}, expressing the Bessel function in terms of the exponential function, and applying the same discussion above, we find: 
%------------------------------------------------------------------------------------------%
\al{
    \xi^{\delta\mC{K}} (\bx,d) 
    &= \frac{\gamma}{\mC{H}}[\Theta(x_{\|})+\Theta(-x_{\|})]
    \frac{3\Omega_{\rm m}H_0^2}{2a(d)} 
    \Int{2}{\bk_{\perp}}{(2\pi)^2} \E^{\iu\bk_{\perp}\cdot\bx_\perp} 
    \exp[-(k_\perp d)^2\sigma^2_\theta] bP_{\rm m}(\eta(d),k_{\perp})
    \,, \label{Eq:xi-deltak-err} \\ 
    \xi^{\mC{K}\mC{K}} (\bx,d) 
    &= \INT{}{r'_1}{}{0}{\infty} w^2(d,r'_1) 
    \frac{9\Omega^2_{\rm m}H_0^4}{4a^2(r'_1)} 
    \Int{2}{\bk_{\perp}}{(2\pi)^2} \E^{\iu\bk_{\perp}\cdot(r_1'\bx_\perp/d)} 
    \exp[-(k_\perp r'_1)^2\sigma^2_\theta]
    P_{\rm m}(\eta(r'_1),k_{\perp})
    \,. \label{Eq:xi-kk-err}
}
%------------------------------------------------------------------------------------------%
The exponential factor is similar to those in redshift measurements where errors in redshift determinations lead to an exponential suppression of Fourier modes along the line-of-sight direction. The maximum wavenumber involved in the observed correlation function is roughly limited by $k_{\|}\alt (1+d\mC{H})/d\sigma_d$ and $k_{\perp}\alt 1/d\sigma_\theta$ except the lensing-lensing correlation function. The lensing-lensing correlation contains smaller scale fluctuations, $k\geq 1/d\sigma_\theta$. 

In the following numerical calculations, we simplify values of $\sigma_d$ and $\sigma_\theta$. For a third generation detector network such as the Einstein Telescope \cite{Maggiore:2019:ET} and Cosmic Explorer \cite{Reitze:2019:CE}, we use the error of the luminosity distance measurement for BNSs estimated in \cite{Camera:2013:GWsiren,Namikawa:2015:gw}. We use the localization error estimate for BNSs at $z=1$ as $\sim 2$\,deg in \cite{Sidery:2014,Namikawa:2015:gw}. Since $\sigma_\theta\propto$\,1/SNR\,$\propto D$ \cite{Cutler:2009:GWcosmo}, we assume $\sigma_\theta=2D/D(z=1)$\,deg. The SNR of BBHs is roughly $10$ times better than those of BNSs \cite{Nishizawa:2019}, and we simply assume that the luminosity distance and localization errors of BBHs are $10\%$ of those of BNSs. We show in Fig.~\ref{fig:kmax} the wavenumbers correspond to $k_{\|}=(1+d\mC{H})/d\sigma_d$ and $k_{\perp}=1/d\sigma_\theta$ as a function of $z$ for BNSs. The limitations to the detectable wavenumbers reduce the signal-to-noise of the multipole power spectrum as we discuss below. 

\subsection{Multipole Power Spectrum}

Next we show the multipole power spectrum. We first define the line-of-sight dependent power spectrum as: 
%------------------------------------------------------------------------------------------%
\al{
    P(k,\mu,d) = \Int{3}{\bx}{} \E^{-\iu\bx\cdot\bk} \xi(\bx,d)
    \,, \label{Eq:xi-to-pk}
}
%------------------------------------------------------------------------------------------%
where $\mu=\hk\cdot\hd$. The multipole power spectrum is defined as a multipole expansion of the above anisotropic power spectrum by the Legendre polynomial, $\mC{P}_\l$: 
%------------------------------------------------------------------------------------------%
\al{
    P_\l(k,d) = \frac{2\l+1}{2}\INT{}{\mu}{}{-1}{1} \mC{P}_\l(\mu) P(k,\mu,d) 
    \,. \label{Eq:pk-to-pkl}
}
%------------------------------------------------------------------------------------------%
The multipole power spectrum is related to the multipole of the correlation function as: 
%------------------------------------------------------------------------------------------%
\al{
    P_\l(k,d) = 4\pi(-\iu)^\l\Int{}{x}{}x^2j_\l(kx)\xi_\l(x,d) 
    \,, \label{Eq:pkl-xil}
}
%------------------------------------------------------------------------------------------%
where we define: 
%------------------------------------------------------------------------------------------%
\al{
    \xi(\bx,d) = \sum_\l \xi_\l(x,d) \mC{P}_\l(\hx\cdot\hd)
    \,. \label{Eq:xil}
}
%------------------------------------------------------------------------------------------%
Note that the lensing term is not simply expanded by the Legendre polynomials as the term is highly anisotropic. We numerically compute the multipole power spectrum from the correlation function using \eqs{Eq:pkl-xil,Eq:xil}. 

In actual observations, the length of the separation is finite. We consider the following window to be multiplied to the correlation function before the integral over the separation vector in \eqs{Eq:pkl-xil}, as similar to \cite{Tansella:2017:RSDfull}: 
%------------------------------------------------------------------------------------------%
\al{
    W(x) = \E^{-(x/x_0)^2} 
    \,,
}
%------------------------------------------------------------------------------------------%
where we define a scale, $x_0$, roughly equivalent to the maximum distance between sources in computing the two-point correlation function. This window function is also important to converge a numerical calculation of the lensing-related multipole power spectra. For the computational convenience of the lensing multipole power spectrum, we adopt $x_0=500$ Mpc. 

%------------------------------------------------------------------------------------------%
\begin{figure}
\centering
\includegraphics[width=140mm]{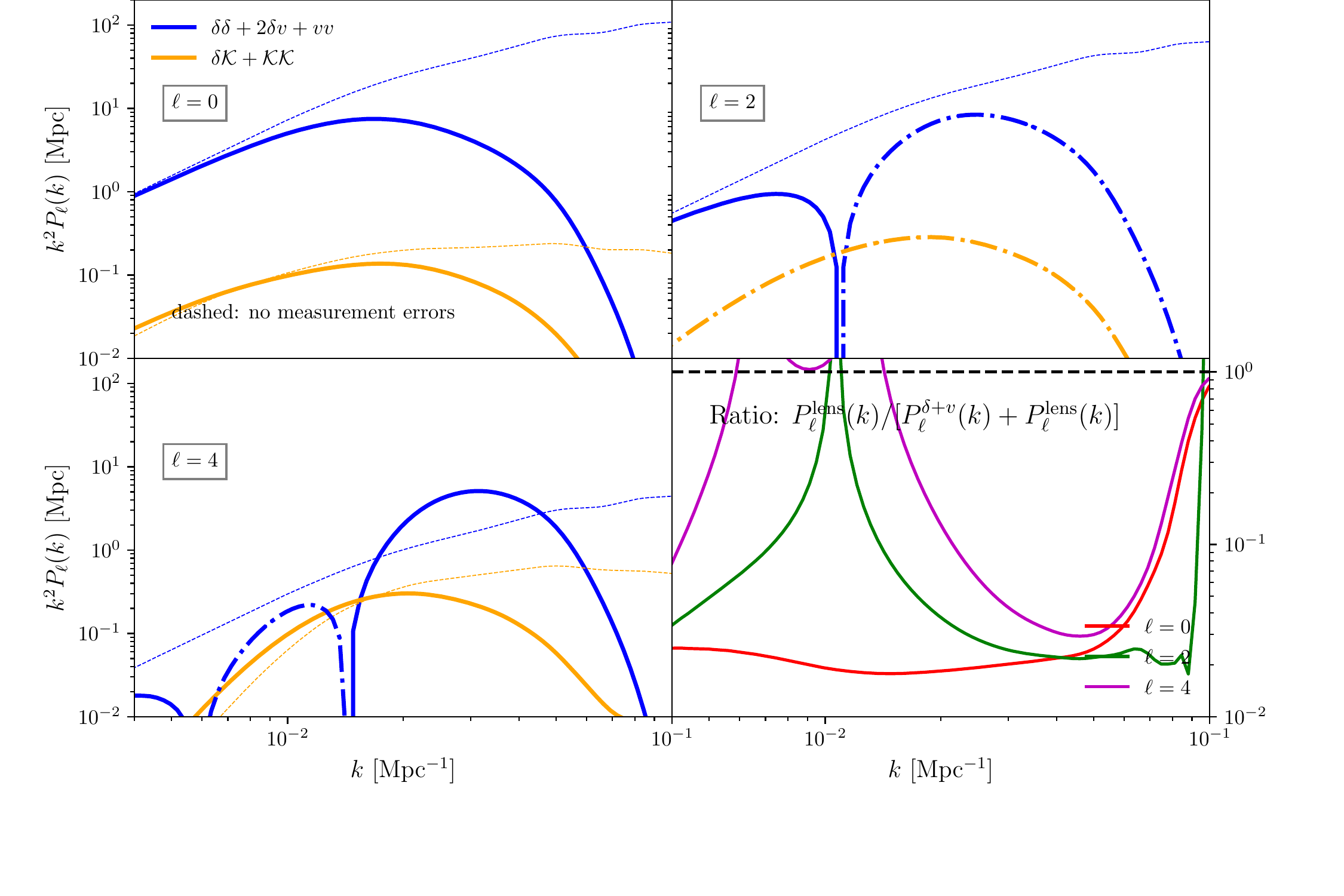}
\caption{
Monopole (Top Left), quadrupole (Top Right) and hexadecapole power spectra (Bottom Left) of BNSs at $D=6.6$\,Gpc (equivalent to $z=1.0$) with the impact of uncertainties in the luminosity distance and angular position in a third generation GW detector network. The dashed lines show the case without the errors and the dot-dashed lines mean negative. The bottom right panel shows the ratio of the lensing contributions ($\delta\mC{K}+\mC{K}\mC{K}$) to the total contribution. 
}
\label{fig:pkl:z1.0:bns}
\end{figure}
%------------------------------------------------------------------------------------------%

%------------------------------------------------------------------------------------------%
\begin{figure}
\centering
\includegraphics[width=140mm]{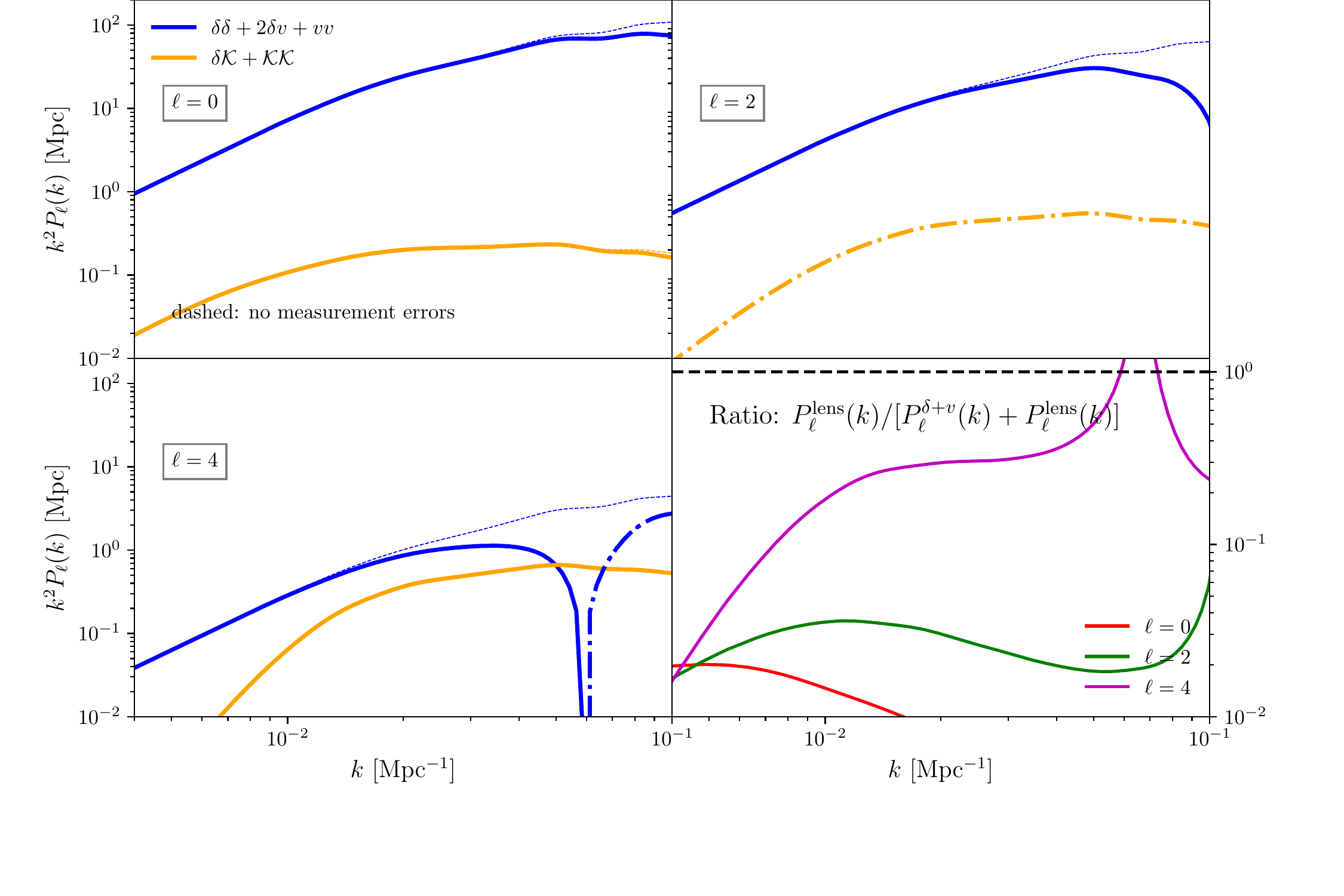}
\caption{
Same as Fig.~\ref{fig:pkl:z1.0:bns} but for BBHs. 
} 
\label{fig:pkl:z1.0:bbh}
\end{figure}
%------------------------------------------------------------------------------------------%

Fig.~\ref{fig:pkl:z1.0:bns} shows the multipole power spectra induced by the density and velocity terms and sum of density-lensing and lensing auto correlations. The lensing only affects the Fourier modes perpendicular to the line-of-sight. The multipole power spectrum is an angle-averaged quantity. Therefore, the lensing effect on the multipole power spectrum is less significant than that on the correlation function. However, the higher-multipole spectra, especially $P_4(k)$, are still affected by the correlation between density and line-of-sight derivative of lensing convergence. The lensing term contributes more significantly to the multipole power spectra than it does in the redshift survey \cite{Tansella:2017:RSDfull}. Fig.~\ref{fig:pkl:z1.0:bbh} plots the case for BBHs, showing that small scale Fourier modes are also detectable due to the high SNR. 

\subsection{Signal-to-Noise of Multipole Power Spectrum}

\begin{figure}
\centering
\includegraphics[width=80mm]{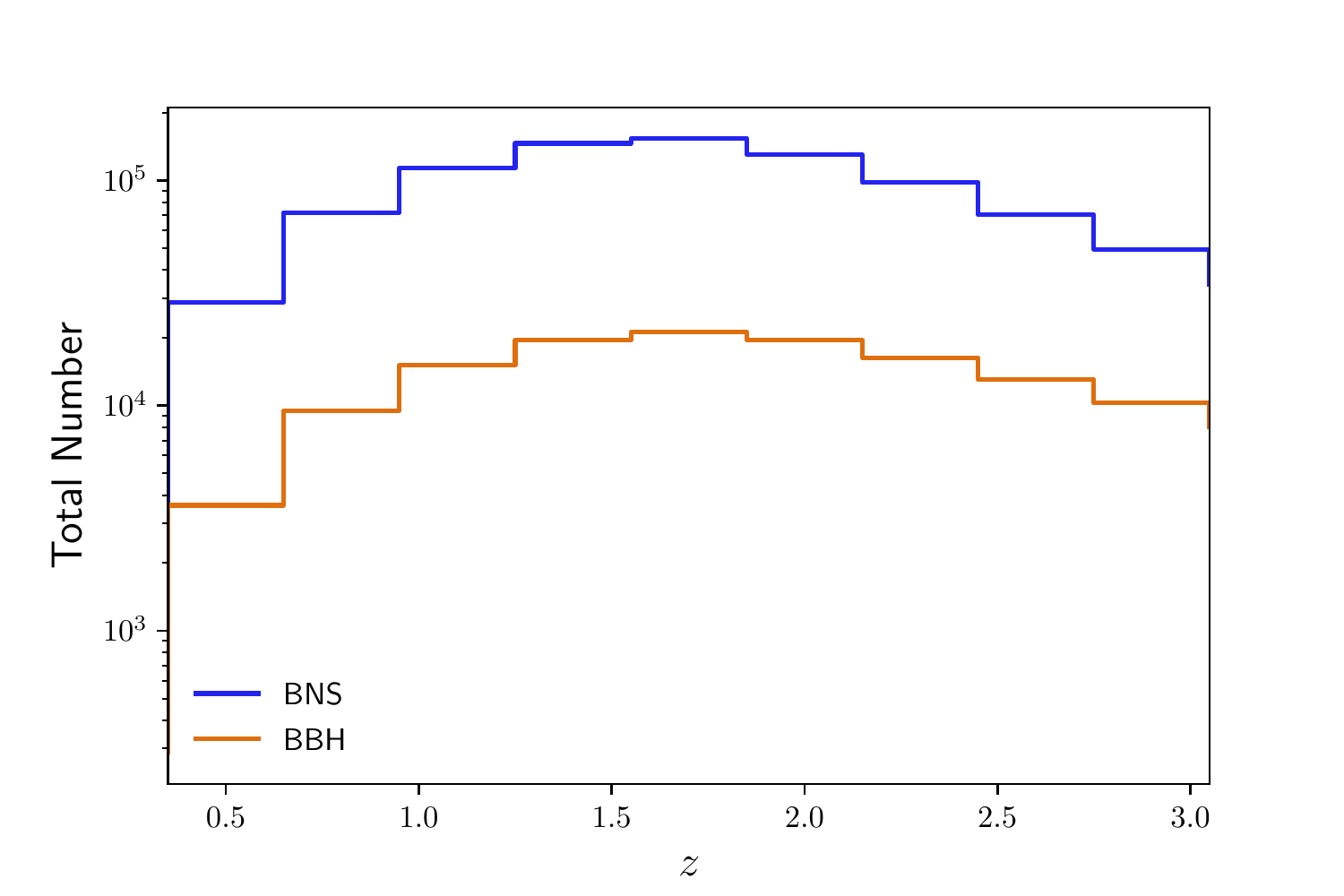}
\caption{
The total number of GW sources at each line-of-sight bin where the bin width is determined so that the corresponding redshift interval is $0.3$. We assume a 3 year observation of a third generation GW detector network.
} 
\label{fig:N}
\end{figure}

This section shows the impact of the luminosity-distance space distortions to the signal-to-noise ratio of the multipole power spectrum. In GW observations, we use the multipole power spectrum estimator developed by \cite{Yamamoto:2005:RSDest} as similar to the galaxy survey. The estimator is defined as \cite{Yamamoto:2005:RSDest}: 
%------------------------------------------------------------------------------------------%
\al{
    \widehat{P}_\l (k)
    &= \frac{2\l+1}{V}\Int{2}{\hk}{4\pi}\INT{3}{\bd}{}{V}{}\INT{3}{\bx}{}{}{}
    \E^{-\iu\bk\cdot\bx} \widehat{\delta}(\bd)\widehat{\delta}(\bd+\bx) \mC{P}_\l(\hk\cdot\hd)
    \,, \label{Eq:est:PP}
}
%------------------------------------------------------------------------------------------%
where $V$ is the survey volume and we apply the uniform weight to the density fluctuations. In the thin shell limit, $V\simeq 4\pi d^2\Delta d$, the expectation value of the above estimator equals to the multipole power spectrum after subtracting the shot noise component \cite{Yamamoto:2005:RSDest,Castorina:2017:RSDest}. The variance of the estimator is given by \cite{Yamamoto:2005:RSDest,Taruya:2010:RSD}: 
%------------------------------------------------------------------------------------------%
\al{
    \ave{|\widehat{P}_\l(k)|^2} 
    &= \frac{4\pi^2 (2\l+1)}{Vk^2\Delta k} \INT{}{\mu}{}{-1}{1}\frac{2\l+1}{2} 
    [\mC{P}_\l(\mu)]^2 \left[P(k,\mu)+\frac{1}{n_{\rm s}(d)}\right]^2
    \,, 
}
%------------------------------------------------------------------------------------------%
where $\Delta k$ is the bin width of $k$ and $n_{\rm s}(d)$ is the source number density at a distance, $d$, defined in \eq{Eq:ns-z}. For computational convenience, in the followings, the line-of-sight bin width is determined so that the corresponding redshift interval is $0.3$. In our setup for the signal-to-noise calculation below, as shown in Fig.~\ref{fig:N}, the total number of events at each line-of-sight bin is at most $10^4$ for BBHs and $10^5$ for BNSs in fullsky which are at least $4$-$5$ orders of magnitude smaller than the number of galaxies observed from recent and future galaxy surveys. The variance of the power spectrum is then dominated by the shot noise and is simplified as: 
%------------------------------------------------------------------------------------------%
\al{
    \ave{|\widehat{P}_\l(k)|^2} 
    &\simeq \frac{4\pi^2 (2\l+1)}{Vk^2\Delta k}\frac{1}{n_{\rm s}^2(d)}
    \,. 
}
%------------------------------------------------------------------------------------------%
The total signal-to-noise of each multipole at each line-of-sight bin is then given by: 
%------------------------------------------------------------------------------------------%
\al{
    \left(\frac{S}{N}\right)^2_{\l,d} = \frac{Vn^2_{\rm s}(d)}{4\pi^2(2\l+1)}
    \INT{}{k}{}{k_{\rm min}}{k_{\rm max}} k^2P_\l^2(k,d)
    \,. 
}
%------------------------------------------------------------------------------------------%
The signal-to-noise is sensitive to a choice of $k_{\rm max}$ but not to $k_{\rm min}$. We set $k_{\rm max}=0.1$\,Mpc$^{-1}$ to focus on the linear scale of the density perturbations. We compute the signal-to-noise at $z\geq0.35$ since at low redshifts the corresponding line-of-sight comoving distance does not satisfy the plane-parallel approximation, $d\alt x_0=500$\,Mpc. 

%------------------------------------------------------------------------------------------%
%------------------------------------------------------------------------------------------%

In our calculation, we assume the duration of observation as $T_{\rm obs}=3$\,yr. For BBHs, \cite{LIGO:2016:bbhrate} derives the local merger rate as $\dot{n}^{\rm BBH}_0=9$-$240$ Gpc$^{-3}$yr$^{-1}$. The local BNS merger rate is estimated by \cite{LIGO:2017:BNS} as $\dot{n}^{\rm BNS}_0=1540^{3200}_{-1220}$ Gpc$^{-3}$yr$^{-1}$. But these estimates still have a large uncertainty \cite{LIGO:2018:O1O2}, and we assume $\dot{n}_0^{\rm BBH}=10^2$ Gpc$^{-3}$yr$^{-1}$ and $\dot{n}^{\rm BNS}_0=10^3$ Gpc$^{-3}$yr$^{-1}$. The shot noise is then obtained from the comoving number density of \eq{Eq:ns-z}. Most of BNSs and BBHs would be detected in the third generation detector networks up to very high redshifts (e.g. \cite{Mills:2017:gw,Scelfo:2018:gw}), and we do not decrease the number of sources by multiplying the detection rate. The following equation shows the scaling of the signal-to-noise with respect to the number of GW sources: 
%------------------------------------------------------------------------------------------%
\al{
    \left(\frac{S}{N}\right)_{\l,d} 
    &=\frac{T_{\rm obs}\dot{n_0}}{3\times 10^2 {\rm Gpc}^{-3}}
    \frac{\mC{R}[z(d)]}{\mC{R}^{\rm fid}[z(d)]}
    \left(\frac{S}{N}\right)^{\rm fid}_{\l,d} 
    \,. 
}
%------------------------------------------------------------------------------------------%

%------------------------------------------------------------------------------------------%
\begin{figure}
\centering
\includegraphics[width=75mm]{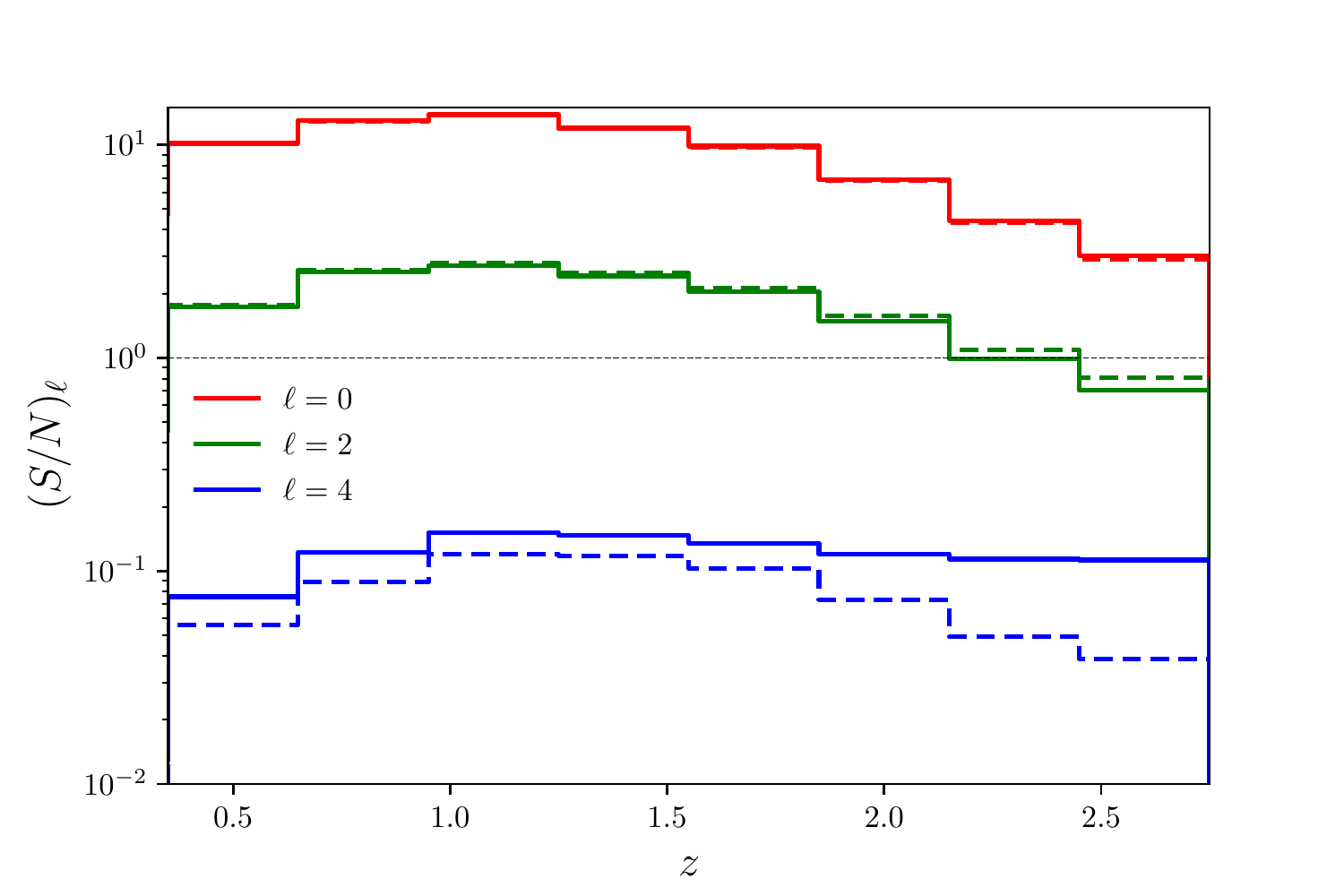}
\includegraphics[width=75mm]{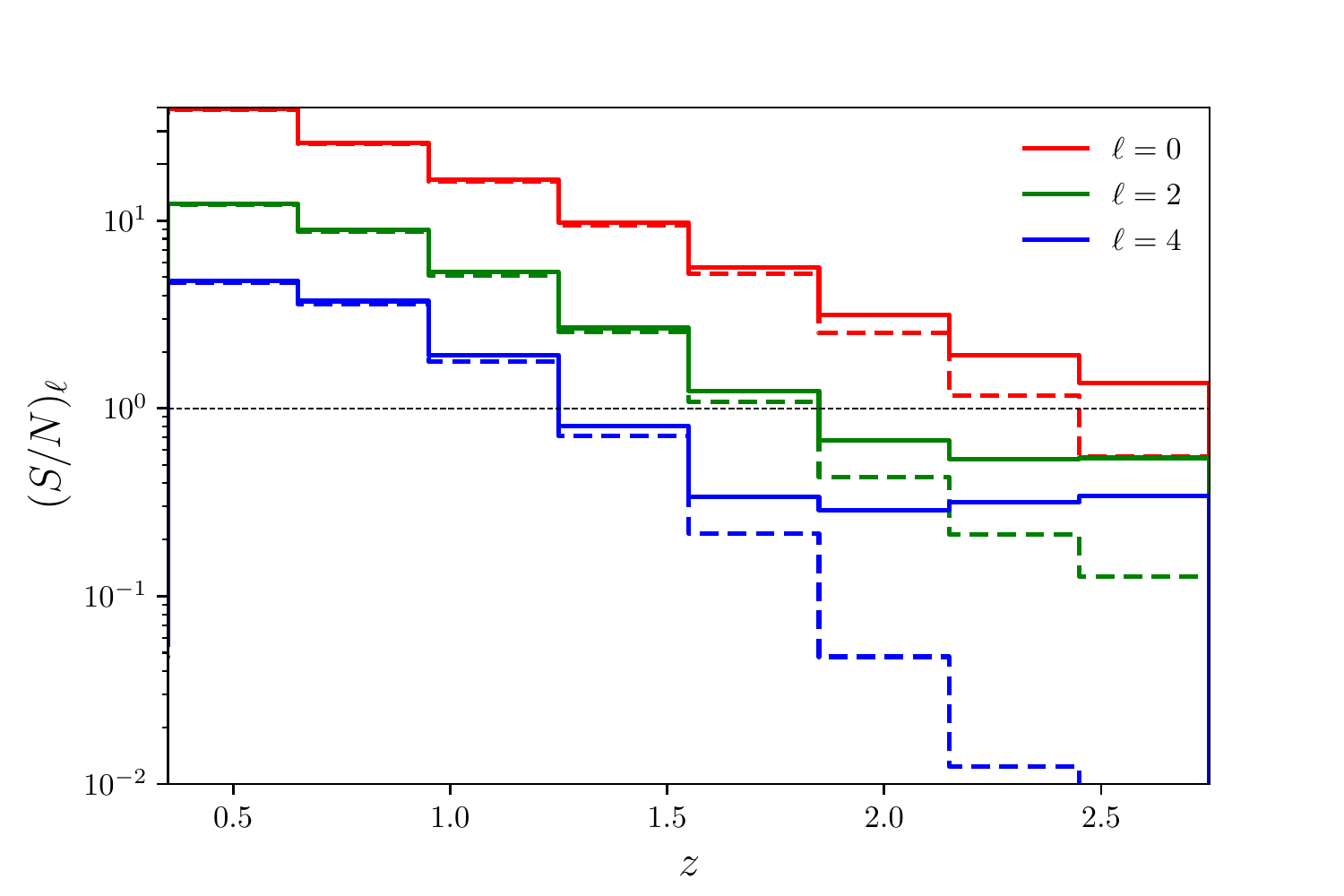}
\caption{
The expected signal-to-noise ratio of the multipole power spectrum at each line-of-sight bin for BBHs (Left) and BNSs (Right) from a third generation detector network with a 3 year observation. The width of the bins corresponds to $\Delta z=0.3$. The solid and dashed lines show the cases including all of the contributions and ignoring the lensing term, respectively.
} 
\label{fig:snr}
\end{figure}
%------------------------------------------------------------------------------------------%

Fig.~\ref{fig:snr} shows the signal-to-noise of the multipole power spectrum as a function of central redshift of the power spectrum measurement. For BBHs, the total signal-to noise of $\l=4$ is also affected by the lensing term. This is because the multipole power spectrum of $\l=4$ has non-negligible contributions from lensing at high $k$ where most of the signal-to-noise comes from, 

\subsection{Cross Correlation between GW Sources and Galaxies}

Since the number density of GW sources is smaller than that of galaxies, the cross correlations between GW sources and galaxies have higher signal-to-noise than the auto correlations discussed in the previous section. Here, we show how the luminosity-distance space distortions affect the cross-correlation function and cross-power spectrum. We define the cross-correlation function as (e.g. \cite{Martinez:1998:cross-xi,Chen:2007:cross-xi,Park:2013:21-gal}): 
%------------------------------------------------------------------------------------------%
\al{
    \xi^{\rm X}(\bx,d) = \ave{\delta_s(\bd)\delta^{\rm gal}_s(\bx+\bd)}
    \,. 
}
%------------------------------------------------------------------------------------------%
From \eqs{Eq:delta-all-kH,Eq:delta-all-kH-gal}, the correlation function is decomposed into:  
%------------------------------------------------------------------------------------------%
\al{
    \xi^{\rm X}(\bx,d) \simeq \xi^{\delta+v,\rm X}(\bx,d) 
    + \frac{b'}{b}\xi^{\delta\mC{K}}(\bx,d) 
    \,, \label{Eq:cross-xi}
}
%------------------------------------------------------------------------------------------%
where $b$ and $b'$ are the linear bias parameters of GW sources and galaxies, respectively, and we define: 
%------------------------------------------------------------------------------------------%
\al{
    \xi^{\delta+v,\rm X}(\bx,d) 
    &\equiv \Int{3}{\bk}{(2\pi)^3} \E^{\iu\bk\cdot\bx}
    \exp\left[-\frac{(k_{\|}d)^2\sigma_d^2}{2(1+d\mC{H})^2}\right]
    \exp\left[-\frac{(k_\perp d)^2\sigma_\theta^2}{2}\right]
    \notag \\
    &\qquad\times
    [bb'+(2\gamma b'+b)\mu^2f+2\gamma\mu^4f^2] P_{\rm m}(\eta(d),k)
    \,. \label{Eq:cross:d-v}
}
%------------------------------------------------------------------------------------------%
Here, we ignore the uncertainty in the redshift, and the uncertainty in $\bx$ only comes from the luminosity distance and localization errors. We use the distant-observer approximation. The correlations between $\delta_{\rm m}$ and $\kappa$, and $\kappa$ and $\mC{K}$, are much smaller than $\xi^{\delta\mC{K}}$ (see Fig.~\ref{fig:xi_x}), and we ignore them in the above equation. The cross-power spectrum and its multipole expansion are given by \eqs{Eq:xi-to-pk,Eq:pk-to-pkl} but with the correlation function described in \eqs{Eq:cross-xi,Eq:cross:d-v}. The estimator is given by \eq{Eq:est:PP} and its variance becomes: 
%------------------------------------------------------------------------------------------%
\al{
    \ave{|\widehat{P}^{\rm X}_\l(k)|^2} 
    &\simeq \frac{2\pi^2 (2\l+1)}{V'k^2\Delta k}\frac{1}{n_{\rm s}(d)}
    \left[\frac{1}{n_{\rm gal}(d)}+{\rm Var}_\l(k)\right]
    \,, 
}
%------------------------------------------------------------------------------------------%
where $V'$ is the overlap volume, $n_{\rm gal}$ is the galaxy number density and the cosmic variance of the galaxy density fluctuations is given by: 
%------------------------------------------------------------------------------------------%
\al{
    {\rm Var}_\l(k) 
    &\equiv \INT{}{\mu}{}{-1}{1}\frac{2\l+1}{2} [\mC{P}_\l(\mu)]^2P^{\rm gal}(k,\mu)
    = (2\l+1)\sum_{\l'=0}^{2\l}\Wjm{\l}{\l}{\l'}{0}{0}{0}^2 P^{\rm gal}_{\l'}(k)
    \,, 
}
%------------------------------------------------------------------------------------------%
Note that: 
%------------------------------------------------------------------------------------------%
\al{
    {\rm Var}_0(k) &= P^{\rm gal}_0(k)
    \,, \\
    {\rm Var}_2(k) &= P^{\rm gal}_0(k) + \frac{2}{7}P^{\rm gal}_2(k) + \frac{2}{7}P^{\rm gal}_4(k)
    \,, \\
    {\rm Var}_4(k) &= P^{\rm gal}_0(k) + \frac{20}{77} P^{\rm gal}_2(k) 
    + \frac{162}{1001}P^{\rm gal}_4(k) + \frac{20}{143} P^{\rm gal}_6(k)
    + \frac{490}{2431}P^{\rm gal}_8(k)
    \,.
}
%------------------------------------------------------------------------------------------%
In our setup, the observed density fluctuations of GW sources are shot-noise dominant as we explained above, and the cross-power spectrum in the variance is negligible.

%------------------------------------------------------------------------------------------%
\begin{figure}
\centering
\includegraphics[width=75mm]{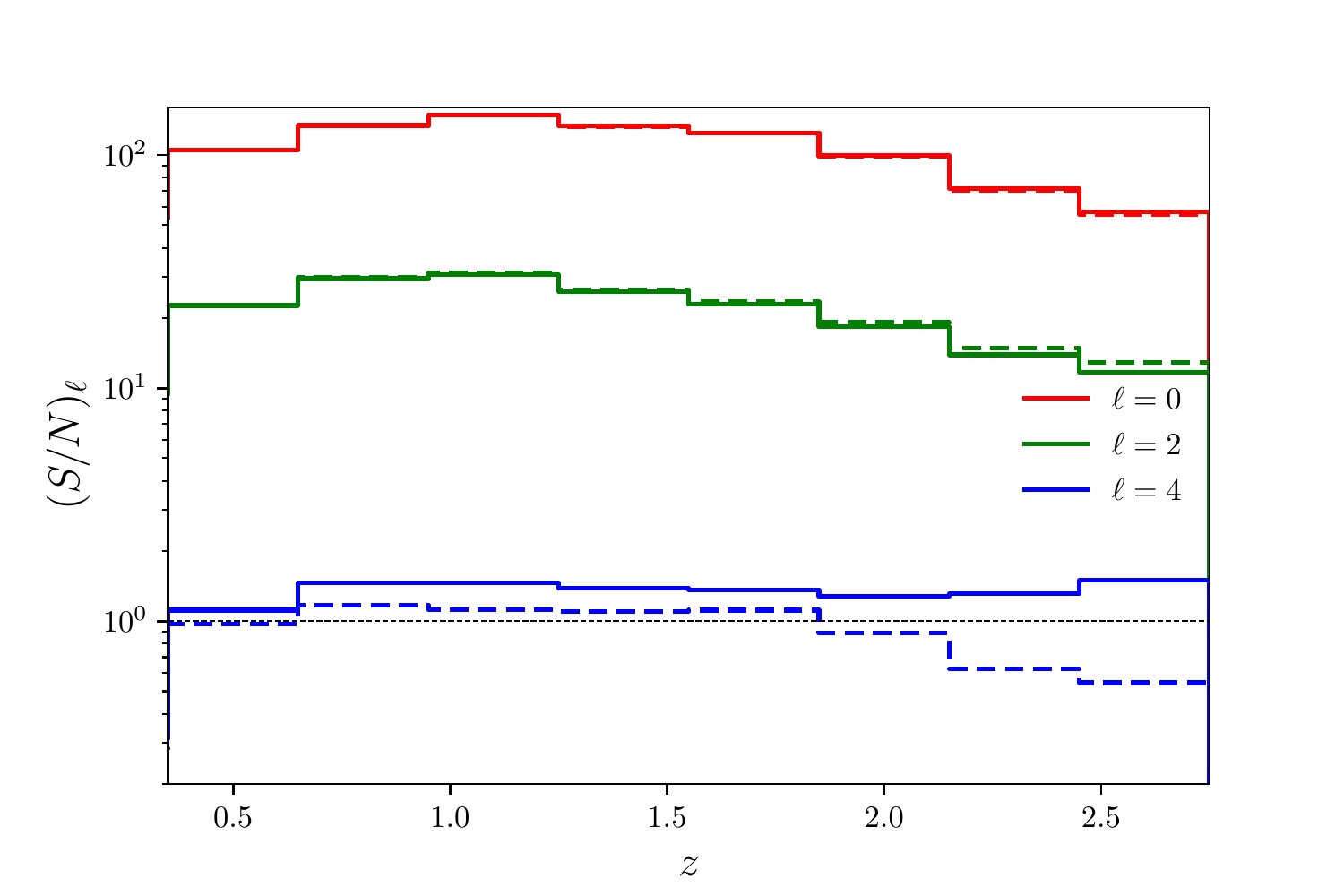}
\includegraphics[width=75mm]{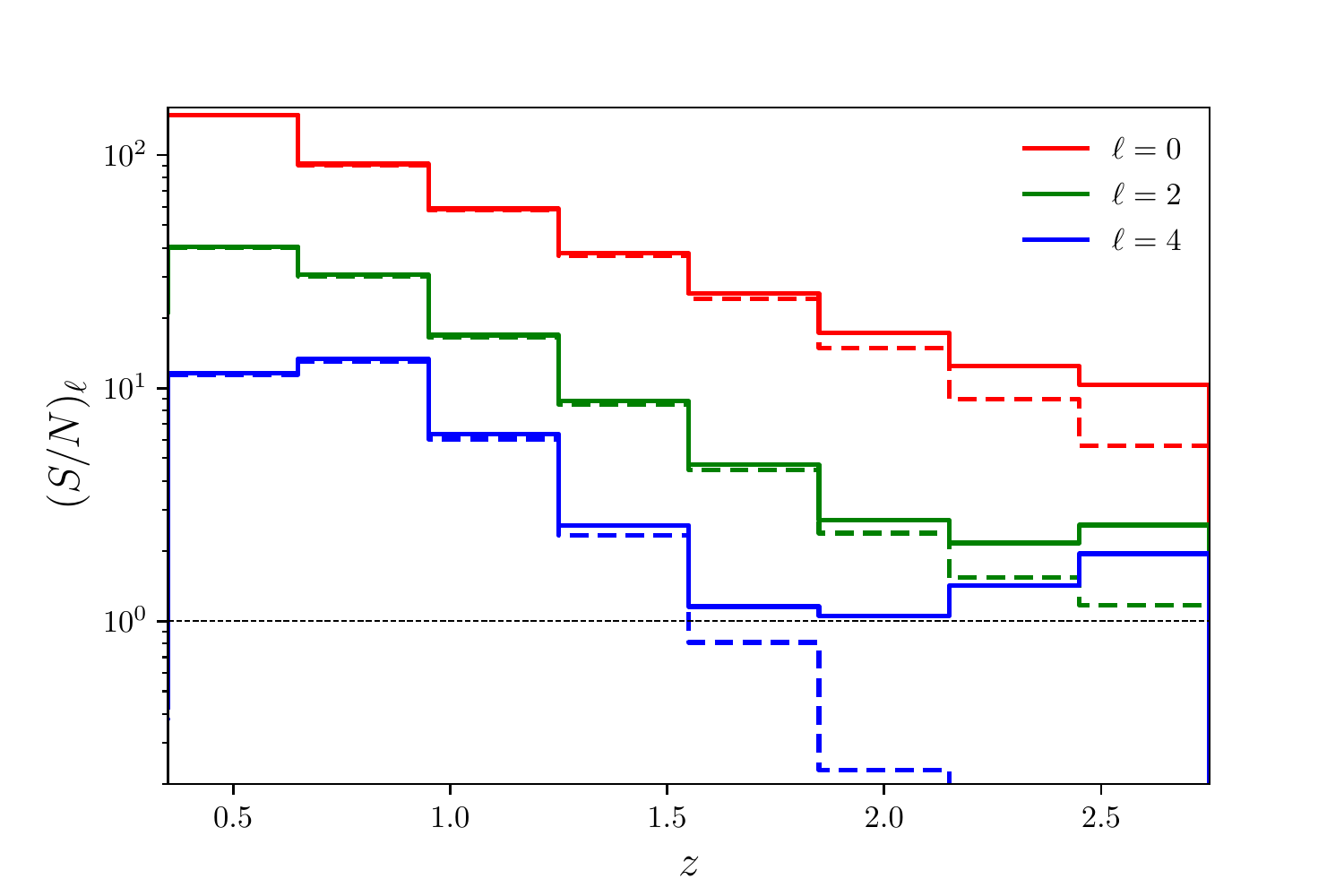}
\caption{Same as Fig.~\ref{fig:snr} but for the cross spectrum between BBHs and galaxies (Left) and between BNSs and galaxies (Right).}
\label{fig:snr:gwxgal}
\end{figure}
%------------------------------------------------------------------------------------------%

Fig.~\ref{fig:snr:gwxgal} shows the signal-to-noise of the cross-power spectrum defined as: 
%------------------------------------------------------------------------------------------%
\al{
    \left(\frac{S}{N}\right)^2_\l = \frac{V'n_{\rm s}(d)n_{\rm gal}(d)}{2\pi^2(2\l+1)}
    \INT{}{k}{}{k_{\rm min}}{k_{\rm max}} 
    k^2 \frac{[P^{\rm X}_\l(k,d)]^2}{1+n_{\rm gal}(d){\rm Var}_\l(k)}
    \,. 
}
%------------------------------------------------------------------------------------------%
For simplicity, we assume a full-sky galaxy observation with a constant galaxy number density $n_{\rm gal}=3\times10^{-2}$\,Mpc$^{-3}$ \cite{Euclid12}, a linear bias $b'=b=\sqrt{1+z}$, no errors in redshift measurements. We ignore the magnification bias since its impact on the power spectrum is small \cite{Hui:2007:RSDmag}. Then, the multipole power spectra, $P_6^{\rm gal}$ and $P_8^{\rm gal}$, vanish. We assume that the line-of-sight bins for GW sources and galaxies are determined so that these two bins are completely overlapped and $\Delta z=0.3$. \eqs{Eq:cross-xi} indicates that the impacts of the lensing term in the luminosity-distance space distortion on $\xi^{\rm X}$ and $P^{\rm X}_\l$ are similar to that on the auto correlation function, $\xi$, and the multipole power spectrum, $P_\l$. However, the total number of observable galaxies is much larger than that of GW sources, and the hexadecapole of the BBHs-galaxy cross-spectrum is detectable. This indicates that we can use this hexadecapole cross-spectrum to extract lensing signals. 

%::::::::::::::::::::::::::::::::::::::::::::::::::::::::::::::::::::::::::::::::::::::::::%
\section{Summary and Discussion} \label{sec:summary}
%::::::::::::::::::::::::::::::::::::::::::::::::::::::::::::::::::::::::::::::::::::::::::%

We derived the density fluctuations obtained from the observed luminosity distance and angular position in GW observations. We found that the density fluctuations in the luminosity-distance space is given by: 
%------------------------------------------------------------------------------------------%
\al{
    \delta_s(\bs) = \Int{3}{\bk}{(2\pi)^3}\E^{\iu\bk\cdot\bs}
    \left(1+2\gamma\beta\mu^2\right) b\delta_{\rm m}(\eta(s),\bk)
    + \mC{K}(s,\hatn)
    \,, 
}
%------------------------------------------------------------------------------------------%
at $k\gg\mC{H}$. The lensing contribution $\mC{K}$ is defined in \eq{Eq:lens-term}. Compared to the redshift-space distortion, there are several differences: 
\bi
\item The dipole term from the velocity perturbations vanishes by ignoring the higher order terms, $\mC{O}(1/k_H)$, not by the distant observer approximation, 
\item The quadrupole term from the velocity perturbations is further multiplied by $2\gamma$ which is consistent with \cite{Zhang:2018:GWPS}, and 
\item The lensing contribution contains ${\rm d}\kappa/{\rm d}r$ in addition to the term proportional to $\kappa$. 
\ei 
We showed that the lensing contributions, in particular, the line-of-sight derivative of the lensing convergence substantially modifies the two-point correlation function at $z\alt 1$. We also showed that the impact of the lensing term on the multipole power spectra are, however, reduced due to the angle average. We also provided a simple estimate of the signal-to-noise ratio of the BBH and BNS multipole power spectra for the third generation detector network. 
The luminosity-distance space distortion is important in the cross-power spectrum with the galaxy number density because the signal-to-noise is high and even the hexadecapole of the BBHs/BNSs-galaxy cross-spectrum would be detectable at high redshifts due to the lensing. 

% lensing effect
Note that we can mitigate the lensing contributions in the power spectrum estimate by removing the Fourier modes perpendicular to the line-of-sight. On the other hand, the lensing is also considered as a signal. To maximally extract information on large-scale structure, we would be able to separately extract the density and velocity terms and lensing signals through the dependence on $k_\|$ and $k_{\perp}$. 

% magnification
In deriving the luminosity-distance space distortions, we have ignored the impact of the perturbations on the number of detectable sources. In galaxy surveys, the lensing magnifies (de-magnifies) sources and fainter (brighter) sources around a detection threshold becomes detectable (undetectable). This modifies the observed number density fluctuations. For a GW observation, the perturbations in the observed luminosity distance modify the SNR of GW events, $\rho$. Denoting the SNR threshold of detection as $\rho_0$, the correction to the observed number density is given by $\delta\rho\times {\rm d}n_s(>\rho_0)/{\rm d}\rho_0$ where $\delta\rho=\rho_0(1-\epsilon)$. The density perturbations then have additional fluctuations, $\delta_{\rm mag}=-s_{\rm gw}\epsilon\sim s_{\rm gw}\kappa$, with $s_{\rm gw}={\rm d}\ln n_s(>\rho_0)/{\rm d}\ln\rho_0$ being a logarithmic slope parameter for GW observations. For BBHs, \cite{Scelfo:2018:gw} estimates the fraction of the detectable sources and a slope parameter, $s'_{\rm gw}=s_{\rm gw}/8\ln 10$. In a third generation GW detectors, they show that the fraction is close to unity for BBHs and the magnification effect would be negligible. The detection rate of BNSs is evaluated in e.g. \cite{Safarzadeh:2019:gw} and would be close to unity for a third generation GW detectors except high redshifts. 

% angular power spectrum
We have focused on the impact of the luminosity-distance space distortions on the correlation function and multipole power spectrum. Alternatively, we can employ the angular power spectrum for detecting the clustering signal of GW sources \cite{Namikawa:2015:gw,Namikawa:2016:gwc,Oguri:2016:GW}. The velocity contribution to the redshift-space angular power spectrum is explored in e.g.~\cite{Padmanabhan:2006:sdss}, showing that the velocity perturbations lead to a modification in the angular power spectrum at $\l\alt 30$. The only difference of the velocity contribution in the luminosity-distance space and redshift space distortions is the factor $2\gamma$. Therefore, the velocity perturbations introduce a modification in the angular power spectrum of the luminosity-distance space at very large scales. The lensing effect is explored in \cite{Namikawa:2015:gw} with a radial distance weight optimized to extract a lensing signal. The impact of lensing is $\sim 10\%$ of the matter density contribution at large scales and is negligible at small scales even at high redshifts. From the above discussion, the luminosity-distance space distortions would be not completely negligible when we analyze the angular power spectrum of the projected density fields on large scales. 

% wide-angle effect
We have computed the correlation function and power spectrum in the plane parallel limit. The plane parallel approximation would be, however, not accurate when we measure the correlation function at a large separation ($x\agt d$) and the multipole power spectrum at large scales ($kd\alt 1$). The impact of the so-called ''wide-angle'' effect has been discussed for galaxy surveys (see e.g.  \cite{Szapudi:2004:tripolar,Papai:2008:RSD,Raccanelli:2010:RSD,Yoo:2013:RSD,Castorina:2017:RSDest,Tansella:2017:RSDfull}). According to e.g. \cite{Yoo:2013:RSD}, the correlation function and multipole power spectra at $z=1$ and $2$ agree with those in the plane-parallel limit at percent level up to the separation of $x\sim1000$~Mpc and $k\sim0.001$~Mpc$^{-1}$, respectively. On the other hand, the wide-angle lensing correction at $z=1$ differs from that in the plane parallel approximation by $20$-$40$\% on large scales ($x\sim 400$~Mpc) \cite{Tansella:2017:RSDfull}. When combining all terms, the correction to the plane parallel approximation at $z\sim 1$ would be smaller than $\sim10$\% even at $x\sim 400$~Mpc. Although the study is based on the redshift space, the impact of the wide-angle effect in the luminosity-distance space would be similar. Note that the impact of the wide-angle effect on the SNR is smaller than that on the correlation function and multipole power spectra because the SNR is almost determined by $k_{\rm max}$. An accurate quantitative estimate of its impact, however, requires a formalism of the full-sky luminosity-distance space distortions, and we leave this study for our future work. 

%//////////////////////////////////////////////////////////////////////////////////////////%
% BACK MATTER 
%//////////////////////////////////////////////////////////////////////////////////////////%

% Acknowledgments %
\acknowledgments
TN thanks Atsushi Nishizawa, Shohei Saga, Blake Sherwin and Atsushi Taruya for helpful discussion. For numerical calculations, this paper used resources of the National Energy Research Scientific Computing Center (NERSC), a U.S. Department of Energy Office of Science User Facility operated under Contract No. DE-AC02-05CH11231.

% Appendix %
\appendix
%\input{app_GR}

%//////////////////////////////////////////////////////////////////////////////////////////%
\section{Luminosity Distance Space Distortions in Linear Perturbation Theory} \label{app:full}
%//////////////////////////////////////////////////////////////////////////////////////////%

In this appendix, we derive the density fluctuations estimated from the observed luminosity distance and angular position in linear theory of General Relativity. We employ the approach developed in \cite{Yoo:2009:GR,Bonvin:2011:GR,Challinor:2011:GR,Jeong:2012} and use part of their results. We assume the following metric perturbations \cite{Bonvin:2011:GR}: 
%------------------------------------------------------------------------------------------%
\al{
    {\rm d}s^2 = -a^2(1+2A){\rm d}\eta^2 - 2a^2B_{|i}{\rm d}\eta{\rm d}x^i
    + a^2(1+2H_L)\bar{g}_{ij}{\rm d}x^i{\rm d}x^j + 2a^2H_{T|ij}
    \,, 
}
%------------------------------------------------------------------------------------------%
where $\bar{g}_{ij}$ is the flat three dimensional spatial metric, and the vertical bar denotes the covariant derivative with respect to $\bar{g}_{ij}$. We ignore vector and tensor perturbations and only include linear order perturbations. 

%::::::::::::::::::::::::::::::::::::::::::::::::::::::::::::::::::::::::::::::::::::::::::%
\subsection{Observables in Luminosity Distance Space} 
%::::::::::::::::::::::::::::::::::::::::::::::::::::::::::::::::::::::::::::::::::::::::::%

\def\volp{\widetilde{V}}
\def\Dobs{\widetilde{D}}
\def\nobs{n^{\rm obs}}
\def\thetaobs{\widetilde{\theta}}
\def\phiobs{\widetilde{\varphi}}
\def\hatnobs{\hs}

In a GW observation, we have the observed source number, ${\rm d}\widetilde{N}$, contained in a small observed volume centered at the observed luminosity distance, $\Dobs$, and direction, $\hatnobs=(\thetaobs,\phiobs)$. Using the observed number density, $\nobs(\Dobs,\hatnobs)$, we obtain: 
%------------------------------------------------------------------------------------------%
\al{
    {\rm d} \widetilde{N} = \nobs (\Dobs,\hatnobs)
    V(\Dobs) \sin\thetaobs {\rm d} \Dobs{\rm d}\thetaobs{\rm d}\phiobs 
    \,, 
}
%------------------------------------------------------------------------------------------%
where we define a volume element inferred from an observer: 
%------------------------------------------------------------------------------------------%
\al{
    V(\Dobs)\sin\thetaobs {\rm d}\Dobs{\rm d}\thetaobs{\rm d}\phiobs  
    \equiv \frac{\widetilde{a}}{1+\widetilde{r}_H}\widetilde{a}^3 
    \widetilde{r}^2\sin\thetaobs {\rm d}\Dobs{\rm d}\thetaobs{\rm d}\phiobs 
    \,. 
}
%------------------------------------------------------------------------------------------%
The quantities, $\widetilde{a}$, $\widetilde{r}$ and $\widetilde{r}_H$, are evaluated using the observed luminosity distance, $\Dobs$. The observed source number is also given by \cite{Weinberg:cosmology,Yoo:2009:GR}: 
%------------------------------------------------------------------------------------------%
\al{
    {\rm d}\widetilde{N} = \npsrc(x) 
    \sqrt{-g}\epsilon_{\mu\nu\rho\sigma}u^\sigma
    \PD{x^\mu}{\Dobs}\PD{x^\nu}{\thetaobs}\PD{x^\rho}{\phiobs}
    {\rm d}\Dobs{\rm d}\thetaobs{\rm d}\phiobs
    \equiv \npsrc(x) \volp(\Dobs,\hatnobs) \sin\thetaobs
    {\rm d}\Dobs{\rm d}\thetaobs{\rm d}\phiobs
    \,, \label{Eq:app:Nobs}
}
%------------------------------------------------------------------------------------------%
where $\npsrc$ is the physical number density of sources, $x^\mu$ is a real source position in 4D spacetime whose spatial components are expressed by a Cartesian coordinate, $\epsilon_{\mu\nu\rho\sigma}$ is the Levi-Civita tensor, and $u^\mu$ is the 4 velocity of sources. The position, $x^\mu$, differs from that inferred from an observer since the inhomogeneities of the universe affect the GW propagation. The observed number density is then given by: 
%------------------------------------------------------------------------------------------%
\al{
    \nobs(\Dobs,\hatnobs) 
    = \npsrc(\Dobs,\hatnobs) \frac{\volp(\Dobs,\hatnobs)}{V(\Dobs)} 
    \,. 
}
%------------------------------------------------------------------------------------------%
To compute the fluctuations of the source number density in observations, we average the above number density by the observed directions for a given observed luminosity distance: 
%------------------------------------------------------------------------------------------%
\al{
    \ave{\nobs}_{\hatn}(\Dobs) \equiv \Int{2}{\hatnobs}{4\pi} \nobs(\Dobs,\hatnobs) 
    \,. 
}
%------------------------------------------------------------------------------------------%
We then define the density fluctuations as: 
%------------------------------------------------------------------------------------------%
\al{
    \nobs(\Dobs,\hatnobs) \equiv \ave{\nobs}_{\hatn}(\Dobs)[1+\delta_s(\Dobs,\hatnobs)] 
    \,. 
}
%------------------------------------------------------------------------------------------%
The perturbation, $\delta_s$, is obtained by evaluating the perturbations of $\npsrc$ and $\volp$ in terms of the observed values. 

%::::::::::::::::::::::::::::::::::::::::::::::::::::::::::::::::::::::::::::::::::::::::::%
\subsection{Perturbations} 
%::::::::::::::::::::::::::::::::::::::::::::::::::::::::::::::::::::::::::::::::::::::::::%

We first consider the perturbations of the volume factor defined in \eq{Eq:app:Nobs}: 
%------------------------------------------------------------------------------------------%
\al{
    \volp(\Dobs,\hatnobs) \sin\thetaobs 
    &= \sqrt{-g}\epsilon_{\mu\nu\rho\sigma}u^\sigma
    \PD{x^\mu}{\Dobs}\PD{x^\nu}{\thetaobs}\PD{x^\rho}{\phiobs}
    \\ 
    &= a^4(1+A+3H_L)\left( \frac{1-A}{a}\epsilon_{\mu\nu\rho 0}
    \PD{x^\mu}{\Dobs}\PD{x^\nu}{\thetaobs}\PD{x^\rho}{\phiobs}
    + \frac{v^i}{a}\epsilon_{\mu\nu\rho i}
    \PD{x^\mu}{\Dobs}\PD{x^\nu}{\thetaobs}\PD{x^\rho}{\phiobs}
    \right)
    \,. \label{Eq:app:vol-terms}
}
%------------------------------------------------------------------------------------------%
Since GWs we consider in this paper propagate along the null geodesics \cite{Issacson:1968:GW1,Issacson:1968:GW2} 
\footnote{See also \url{http://www.pmaweb.caltech.edu/Courses/ph136/yr2012/1227.1.K.pdf}}, we define perturbations to coordinate using the unperturbed path $\ol{x}^\mu(\lambda)$ as $\delta x^\mu(\lambda)\equiv x^\mu(\lambda)-\ol{x}^\mu(\lambda)$ where $\lambda$ is the affine parameter of null geodesic \cite{Bonvin:2011:GR}. 
\footnote{Alternatively, we can define the perturbed coordinates using the unperturbed quantities evaluated by the observed redshift \cite{Yoo:2014:GR} to skip discussion around \eq{Eq:app:vol-back}.} 
The first term in the parenthesis is the Jacobian of the spatial coordinate transform which leads to: 
%------------------------------------------------------------------------------------------%
\al{
    \epsilon_{\mu\nu\rho 0} \PD{x^\mu}{\Dobs}\PD{x^\nu}{\thetaobs}\PD{x^\rho}{\phiobs} 
    &= \D{(r+\delta r)}{\Dobs}r^2\left(1+2\frac{\delta r}{r}\right)\sin\thetaobs
    \left[1+\left(\cot\thetaobs+\PD{}{\thetaobs}\right)\delta\theta
    +\PD{}{\phiobs}\delta\varphi\right]
    \,, \label{Eq:app:vol-term1}
}
%------------------------------------------------------------------------------------------%
where $\delta r$, $\delta\theta$ and $\delta\varphi$ are the perturbed line-of-sight comoving distance and angular position of $\delta x^\mu$. The second term in the parenthesis becomes: 
%------------------------------------------------------------------------------------------%
\al{
    v^i\epsilon_{\mu\nu\rho i}
    \PD{x^\mu}{\Dobs}\PD{x^\nu}{\thetaobs}\PD{x^\rho}{\phiobs}
    &= v_r\D{r}{D}r^2\sin\thetaobs
    \,, \label{Eq:app:vol-term2}
}
%------------------------------------------------------------------------------------------%
where $v_r$ is the radial velocity from observer to sources. Substituting \eqs{Eq:app:vol-term1,Eq:app:vol-term2} into \eq{Eq:app:vol-terms}, we obtain up to linear order of perturbations as (see e.g. Eq.~(12) of \cite{Bonvin:2011:GR} for the observed redshift case): 
%------------------------------------------------------------------------------------------%
\al{
    \volp(\Dobs,\hatnobs) 
    &= a^3(1+A+3H_L)\D{r}{D} r^2 \left[
    \D{D}{r}\D{(r+\delta r)}{\Dobs} 
    \left(1+2\frac{\delta r}{r}\right) (1-2\tilde{\kappa}) - A + v_r \right]
    \\ 
    &= V(D)(1+A+3H_L) \left[
    \D{D}{r}\D{(r+\delta r)}{\Dobs}\left(1+2\frac{\delta r}{r}-2\tilde{\kappa}\right)
    -A + v_r \right]
    \,, \label{Eq:app:vol-expand1}
}
%------------------------------------------------------------------------------------------%
where we define: 
%------------------------------------------------------------------------------------------%
\al{
    -2\tilde{\kappa} \equiv \left(\cot\thetaobs+\PD{}{\thetaobs}\right)\delta\theta
    +\PD{}{\phiobs}\delta\varphi
    \,. 
}
%------------------------------------------------------------------------------------------%
Note that, for the first order quantity as a function of time and spatial position, we set the total derivative, ${\rm d}/{\rm d}r=-{\rm d}/{\rm d}\lambda$, and: 
%------------------------------------------------------------------------------------------%
\al{
    \D{r}{\Dobs} 
    &= \left(\D{D}{r}+\D{\delta D}{r}\right)^{-1}
    \simeq \D{r}{D}\left(1+\D{r}{D}\D{\delta D}{\lambda}\right)
    \,, \\ 
    \D{D}{r}\D{(r+\delta r)}{\Dobs} 
    &= \D{D}{r}\left(\D{r}{\Dobs}+\D{\delta r}{D}\right)
    \simeq 1-\D{\delta r}{\lambda} + \D{\delta D}{\lambda}\D{r}{D}
    \,. \label{Eq:app:dDdr2}
}
%------------------------------------------------------------------------------------------%
Substituting \eq{Eq:app:dDdr2} into \eq{Eq:app:vol-expand1}, we obtain: 
%------------------------------------------------------------------------------------------%
\al{
    \volp(\Dobs,\hatnobs) 
    &= V(D) (1+A+3H_L) \left[\left(1-\D{\delta r}{\lambda}
    +\D{\delta D}{\lambda}\D{r}{D}\right) 
    \left(1+2\frac{\delta r}{r}-2\tilde{\kappa}\right) - A + v_r\right]
    \\ 
    &= V(D)\left(1+3H_L + v_r -2\tilde{\kappa} + 2\frac{\delta r}{r}-\D{\delta r}{\lambda} 
    + \D{r}{D}\D{\delta D}{\lambda}\right)
    \,. \label{Eq:app:vol-expand2}
}
%------------------------------------------------------------------------------------------%
To evaluate $V$ at an observed luminosity distance, we use the following relation: 
%------------------------------------------------------------------------------------------%
\al{
    V(\Dobs) = V(D) + \D{V}{D}\delta D
    \,. \label{Eq:app:vol-back}
}
%------------------------------------------------------------------------------------------%
Then, \eq{Eq:app:vol-expand2} is recast as: 
%------------------------------------------------------------------------------------------%
\al{
    \delta_V \equiv \frac{\volp(\Dobs,\hatnobs)}{V(\Dobs)} - 1 
    = 3H_L + v_r -2\tilde{\kappa} + 2\frac{\delta r}{r} - \D{\delta r}{\lambda} 
    + \D{r}{D}\D{\delta D}{\lambda} - \D{\ln V}{D}\delta D
    \,. \label{Eq:app:vol-expand3}
}
%------------------------------------------------------------------------------------------%

Note that \eq{Eq:app:vol-expand3} is gauge-invariant. To see this, we use the fact that the luminosity distance perturbation as a function of the perturbed (observed) redshift, $\epsilon'$, is known as a gauge invariant quantity \cite{Sasaki:1987}. $\epsilon'$ is related to $\epsilon$ as \cite{Bonvin:2005:LD}: 
%------------------------------------------------------------------------------------------%
\al{
    \epsilon = \epsilon'+\frac{1}{\gamma}\delta z'
    \,, \label{Eq:app:epsilon-z}
}
%------------------------------------------------------------------------------------------%
where $\delta z'$ is the redshift perturbation defined as $1+\widetilde{z}=(1+z)(1+\delta z)$ \cite{Yoo:2014:GR}. We rewrite \eq{Eq:app:vol-expand3} using $\epsilon'$ and the volume perturbation in redshift surveys given in Eq.~(17) of \cite{Bonvin:2011:GR}, both of which are gauge invariant \cite{Sasaki:1987,Yoo:2014:GR}. We denote the redshift perturbation of \cite{Bonvin:2011:GR} as $\delta z=\widetilde{z}-z$, and use the following equations: 
%------------------------------------------------------------------------------------------%
\al{
    V_z &\equiv \frac{a^3r^2}{H} 
    \,, \\
    \D{r}{D}\D{\delta D}{\lambda} 
    &= \frac{\gamma}{\mC{H}}\D{\epsilon'}{\lambda}
    - \D{\gamma^{-1}}{D}D\delta z'
    + \frac{1}{\mC{H}}\D{\delta z'}{\lambda}
    - \epsilon'-\frac{1}{\gamma}\delta z'
    \,, \label{Eq:app:drdd} \\
    -\D{\ln V}{D}\delta D
    &= -\D{\ln V}{\ln D}\epsilon' + \D{\gamma^{-1}}{D}D\delta z' 
    - \D{\ln V_z}{z}\delta z + \frac{1}{\mC{H}D}\delta z
    \label{Eq:app:dvdd} \\ 
    \frac{1}{\mC{H}}\D{\delta z'}{\lambda}
    &= \frac{1}{H}\D{\delta z}{\lambda} + \frac{\delta z}{1+z}
    \\
    -\frac{a}{\gamma} + \frac{1}{\mC{H}D} 
    &= -\frac{1}{1+z}
    \,, 
}
%------------------------------------------------------------------------------------------%
Combining \eqs{Eq:app:drdd,Eq:app:dvdd}, we obtain: 
%------------------------------------------------------------------------------------------%
\al{
    \D{r}{D}\D{\delta D}{\lambda} - \D{\ln V}{D}\delta D 
    &= \frac{\gamma}{\mC{H}}\D{\epsilon'}{\lambda} - \epsilon'
    - \D{\ln V}{\ln D}\epsilon' + \frac{1}{H}\D{\delta z}{\lambda}
    - \D{\ln V_z}{z}\delta z
    \,. 
}
%------------------------------------------------------------------------------------------%
Substituting the above equation into \eq{Eq:app:vol-expand3}, we find: 
%------------------------------------------------------------------------------------------%
\al{
    \delta_V 
    = 3H_L + v_r -2\tilde{\kappa} + 2\frac{\delta r}{r} - \D{\delta r}{\lambda} 
    + \frac{1}{H}\D{\delta z}{\lambda} - \D{\ln V_z}{z}\delta z
    + \left(\frac{\gamma}{\mC{H}}\D{}{\lambda} - 1 - \D{\ln V}{\ln D}\right)\epsilon'
    \,. \label{Eq:app:vol-expand4}
}
%------------------------------------------------------------------------------------------%
The above perturbation is the sum of the volume perturbation given by Eq.~(17) of \cite{Bonvin:2011:GR} and $\epsilon'$. Therefore, $\delta_V$ is also gauge-invariant. 

Next we consider the perturbations of the physical source number density. Similar to \cite{Yoo:2014:GR}, the physical source number density is expanded up to first order of perturbations as: 
%------------------------------------------------------------------------------------------%
\al{
    \npsrc(D,\hatn) 
    &= \npsrc(D)[1+\dsrc(D,\hatn)] = \npsrc(\Dobs-\delta D)[1+\dsrc(\Dobs,\hatnobs)]
    \\ 
    &= \npsrc(\Dobs) \left[1 + \dsrc(\Dobs,\hatnobs) 
    - \D{\ln\npsrc}{D}D\epsilon \right]
    \\ 
    &= \npsrc(\Dobs) \left[1 + \dsrc(\Dobs,\hatnobs) 
    - \frac{3}{1+z}\D{z}{D} D\epsilon - \D{\ln(a^3\npsrc)}{D} D\epsilon \right]
    \,. \label{Eq:app:ns-perturb1} 
}
%------------------------------------------------------------------------------------------%
If the source number density is proportional to the matter density, the last term of \eq{Eq:app:ns-perturb1} vanishes. The above equation contains terms due to evaluating fluctuations using observed quantities instead of unperturbed quantities. Note that, in the followings, we use the comoving source number density, $\ncsrc\equiv a^3\npsrc$, and: 
%------------------------------------------------------------------------------------------%
\al{
    \D{z}{D} &= \frac{\mC{H}}{1+r_H} = \frac{\gamma}{r} 
    \,, \\
    \D{\ln\ncsrc}{\ln D} &= \frac{1}{1+r_H}r\D{\ln\ncsrc}{\ln r}
    \,. 
}
%------------------------------------------------------------------------------------------%
\eq{Eq:app:ns-perturb1} is then recast as: 
%------------------------------------------------------------------------------------------%
\al{
    \frac{\npsrc(\Dobs,\hatnobs)}{\npsrc(\Dobs)} 
    =  1 + \dsrc - 3\gamma\epsilon - \D{\ln \ncsrc}{\ln r} \frac{\epsilon}{1+r_H}
    \,. \label{Eq:app:ns-perturb2}
}
%------------------------------------------------------------------------------------------%

%------------------------------------------------------------------------------------------%
\subsection{Newtonian Gauge and Sub-Horizon Limit} 
%------------------------------------------------------------------------------------------%

To obtain a simplified form, we apply the Newtonian gauge; $A=\Psi$, $B=0$, $H_L=-\Phi$ and $H_T=0$. We also ignore the constant and dipole terms since they only affect anisotropies at the largest scales ($\l\leq1$). Under this approximation, $\tilde{\kappa}=\kappa$ which is given by \eq{Eq:kappa} (but replacing the integration variable with $\lambda$). 

We first rewrite \eq{Eq:app:vol-expand3} as: 
%------------------------------------------------------------------------------------------%
\al{
    \delta_V
    &= - \D{\delta r}{\lambda}-\epsilon-\frac{\gamma}{\mC{H}}\D{\epsilon}{r} 
    + 2\frac{\delta r}{r} -2\kappa - v_r -3\Phi + \gamma\epsilon 
    \left(4-\frac{2}{r_H} +\frac{\gamma}{r_H}-\gamma\frac{\mC{H}_{,\eta}}{\mC{H}^2}\right)
    \,, \label{Eq:app:vol-expand5}
}
%------------------------------------------------------------------------------------------%
where we use the derivative of the unperturbed volume element with respect to the luminosity distance: 
%------------------------------------------------------------------------------------------%
\al{
    \D{V(D)}{D} 
    &= -\frac{\gamma}{D}V(D) \left( 4 - \frac{2}{r_H}
    + \frac{\gamma}{r_H} - \gamma\frac{\mC{H}_{,\eta}}{\mC{H}^2} \right)
    \,. 
}
%------------------------------------------------------------------------------------------%
We substitute the solutions of the geodesic equation, $\delta r$ and ${\rm d}\delta r/{\rm d}\lambda$, derived in \cite{Bonvin:2011:GR} to \eq{Eq:app:vol-expand5}, and obtain: 
%------------------------------------------------------------------------------------------%
\al{
    \delta_V
    = \Phi + \Psi - \epsilon - \frac{\gamma}{\mC{H}}\D{\epsilon}{r} 
    + \frac{2}{r}\INT{}{r'}{}{0}{r}(\Phi+\Psi) - 2\kappa - v_r -3\Phi
    + \gamma\epsilon \left(4-\frac{2}{r_H}
    + \frac{\gamma}{r_H}-\gamma\frac{\mC{H}_{,\eta}}{\mC{H}^2}\right)
    \,. 
}
%------------------------------------------------------------------------------------------%
Combining the above equation and \eq{Eq:app:ns-perturb2}, the total perturbations to the observed number density become: 
%------------------------------------------------------------------------------------------%
\al{
    \delta_s 
    &= \dsrc - 2\Phi + \Psi + \frac{2}{r}\INT{}{r'}{}{0}{r}(\Phi+\Psi) 
    - 2\kappa - v_r 
    - \frac{\gamma}{\mC{H}}\D{\epsilon}{r} + \frac{\epsilon}{1+r_H} 
    \left(-3-\D{\ln\ncsrc}{\ln r}+\gamma-\frac{\gamma r\mC{H}_{,\eta}}{\mC{H}}\right)
    \\ 
    &= \dsrc - 2\Phi + \Psi - 2\kappa - v_r
    + \frac{2}{r}\INT{}{r'}{}{0}{r}(\Phi+\Psi) 
    - \frac{\gamma}{\mC{H}}\D{\epsilon}{r} - \alpha\epsilon
    \,. 
}
%------------------------------------------------------------------------------------------%
The last two terms are the same as that derived in the main text. 

The potential terms are negligible at the sub-horizon scale and we obtain: 
%------------------------------------------------------------------------------------------%
\al{
    \delta_s \simeq b\delta_{\rm m} - 2\kappa - v_r
    - \frac{\gamma}{\mC{H}}\D{\epsilon}{r} - \alpha\epsilon
    \,. \label{Eq:app:fluc}
}
%------------------------------------------------------------------------------------------%
Here, we are interested in the sub-horizon scale and the source number density is replaced by $b\delta_{\rm m}$ which is the matter density fluctuations in the synchronous gauge multiplied by a constant bias \cite{Challinor:2011:GR}. The luminosity distance perturbations in the Newtonian gauge as a function of observed redshift are given by (see e.g. Eq.~(3.10) of \cite{Yoo:2016:LD}): 
%------------------------------------------------------------------------------------------%
\al{
    \epsilon' = \left(1-\frac{1}{r_H}\right) \delta z'
    + \frac{1}{r}\INT{}{r'}{}{0}{r}(\Psi+\Phi) - \kappa - \Phi
    \,, \label{Eq:app:epsilon-GI}
}
%------------------------------------------------------------------------------------------%
where the redshift perturbation is given as (see e.g. Eq.~(2.34) of \cite{Yoo:2016:LD}): 
%------------------------------------------------------------------------------------------%
\al{
    \delta z' = -\Psi + v_r - \INT{}{r'}{}{0}{r}(\dot{\Psi}+\dot{\Phi})
    \,. \label{Eq:app:delta-zp}
}
%------------------------------------------------------------------------------------------%
Substituting \eqs{Eq:app:epsilon-GI,Eq:app:delta-zp} into \eq{Eq:app:epsilon-z}, we obtain: 
%------------------------------------------------------------------------------------------%
\al{
    \epsilon = 2v_r - \kappa - \Phi - 2\Psi - 2\INT{}{r'}{}{0}{r}(\dot{\Psi}+\dot{\Phi})
    + \frac{1}{r}\INT{}{r'}{}{0}{r}(\Psi+\Phi)
    \,, \label{Eq:app:delta-DL}
}
%------------------------------------------------------------------------------------------%
Substituting the above equation to \eq{Eq:app:fluc}, and ignoring the potential terms, we find: 
%------------------------------------------------------------------------------------------%
\al{
    \delta_s 
    &\simeq b\delta_{\rm m} - 2\kappa - v_r
    - \frac{\gamma}{\mC{H}}\D{(2v_r-\kappa)}{r} - \alpha (2v_r-\kappa)
    \\ 
    &= b\delta_{\rm m} - (2-\alpha)\kappa - (1+2\alpha)v_r  
    - \frac{2\gamma}{\mC{H}}\D{v_r}{r} + \frac{\gamma}{\mC{H}}\D{\kappa}{r}
    \,, 
}
%------------------------------------------------------------------------------------------%
The term, $(1+2\alpha)v_r$, and conformal time derivative $\pd v_r/\pd\eta$ are negligible in the sub horizon, and the dominant term is given by: 
%------------------------------------------------------------------------------------------%
\al{
    \delta_s \simeq b\delta_{\rm m} - \frac{2\gamma}{\mC{H}}\PD{v_r}{r} 
    + \left(-2+\alpha+\frac{\gamma}{\mC{H}}\D{}{r}\right)\kappa
    \,. 
}
%------------------------------------------------------------------------------------------%
The above equation is equivalent to \eq{Eq:delta-s-lead}.

%\input{app_error}

%//////////////////////////////////////////////////////////////////////////////////////////%
\section{Impact of Errors in Luminosity Distance and Localization on Detectable Fourier Modes} \label{app:error}
%//////////////////////////////////////////////////////////////////////////////////////////%

Here, we derive \eqs{Eq:error:d-v,Eq:xi-deltak-err,Eq:xi-kk-err}. We first assume that the errors in the luminosity distance and sky position are described as a random Gaussian fields and are not correlated between different sources. Denoting the errors in the luminosity distance and $D\to D[1+\epsilon_d(D)]$ and angular position as $\hs\to\hs+\delta\hs$, respectively, the comoving separation vector has the following error: 
%------------------------------------------------------------------------------------------%
\al{
    \delta\bx \equiv \delta\bs_2 - \delta\bs_1 
    \simeq \hs_2 \frac{\gamma_2}{\mC{H}_2}\epsilon_d(D_2)
    - \hs_1 \frac{\gamma_1}{\mC{H}_1}\epsilon_d(D_1)
    + s_2\delta\hs_2 - s_1\delta\hs_1
    \,, 
}
%------------------------------------------------------------------------------------------%
where we only consider the terms up to first order of $\epsilon_d$ and $\delta\hs$. The exponential factor in the first term of \eq{Eq:xi-P} is then replaced by: 
%------------------------------------------------------------------------------------------%
\al{
    \E^{\iu\bk\cdot\bx} \to \E^{\iu\bk\cdot\bx} \ave{\E^{\iu\bk\cdot\delta\bx}}
    \,, 
}
%------------------------------------------------------------------------------------------%
Since $\epsilon_d$ and $\delta n_i$ are zero-mean Gaussian random variables, we obtain: 
%------------------------------------------------------------------------------------------%
\al{
    \ave{\delta\bx} &= 0 
    \,, \\ 
    \ave{\E^{\iu\bk\cdot\delta\bx}} &= \E^{-\ave{[\bk\cdot\delta\bx]^2}/2}
    \,. 
}
%------------------------------------------------------------------------------------------%
In the plane parallel approximation, i.e., $\hs_1\simeq\hs_2\simeq\hd$, $s_1\simeq s_2\simeq d$, we find: 
%------------------------------------------------------------------------------------------%
\al{
    \ave{[\bk\cdot\delta\bx]^2} 
    &\simeq (kd\mu)^2 \frac{\gamma^2}{d^2\mC{H}^2}
    2\AVE{\epsilon^2_d(D)} + k^2(1-\mu^2)d^2 2\ave{|\delta n|^2}
    \\ 
    &= \frac{(kd\mu)^2}{(1+d\mC{H})^2}2\sigma_d^2(D) 
    + k^2(1-\mu^2)d^2 2\sigma^2_\theta(D)
    \,. 
}
%------------------------------------------------------------------------------------------%
The correlation function is then recast as \eq{Eq:error:d-v}. 

For lensing, we first rewrite \eq{Eq:xi-kk} as: 
%------------------------------------------------------------------------------------------%
\al{
    \xi^{\delta\mC{K}} (\bx,d)
    &= \frac{\gamma}{\mC{H}}
    [\Theta(x_{\|})+\Theta(-x_{\|})]
    \frac{3\Omega_{\rm m}H_0^2}{2a(d)} 
    \Int{2}{\bk_{\perp}}{(2\pi)^2} \E^{\iu\bk_{\perp}\cdot\bx_{\perp}}
    bP_{\rm m}(\eta(d),k_{\perp})
    \\ 
    \xi^{\mC{K}\mC{K}} (\bx,d) 
    &= \INT{}{r'}{}{0}{\infty} w^2(d,r') 
    \frac{9\Omega^2_{\rm m}H_0^4}{4a^2(r')} 
    \Int{2}{\bk_{\perp}}{(2\pi)^2} \E^{\iu\bk_{\perp}\cdot\bx_{\perp}(r'/d)}
    P_{\rm m}(\eta(r'),k_{\perp})
    \,. 
}
%------------------------------------------------------------------------------------------%
Here, we use: 
%------------------------------------------------------------------------------------------%
\al{
    J_0(x) = \frac{1}{2\pi}\INT{}{\theta}{}{0}{2\pi}\E^{\iu x\cos\theta}
    \,. 
}
%------------------------------------------------------------------------------------------%
In the presence of errors, the exponential factors in the above equations become: 
%------------------------------------------------------------------------------------------%
\al{
    \ave{\E^{\iu\bk_{\perp}\cdot(\bx_{\perp}+\delta\bx_\perp)}}
    &= \E^{\iu\bk_{\perp}\cdot\bx_\perp} \E^{-(k_\perp d)^2\sigma^2_\theta}
    \,, 
    \\ 
    \ave{\E^{\iu\bk_{\perp}\cdot(\bx_{\perp}+\delta\bx_\perp)(r'/d)}}
    &= \E^{\iu\bk_{\perp}\cdot\bx_\perp(r'/d)} \E^{-(k_\perp r')^2\sigma^2_\theta}
    \,, 
}
%------------------------------------------------------------------------------------------%
where we use $\delta\bx_\perp=d\delta\hs_2-d\delta\hs_1$. The errors also modify $x_{\|}\to x_{\|}+\delta x_{\|}$ but the two-times derivative of the step function becomes zero. We then obtain \eqs{Eq:xi-deltak-err,Eq:xi-kk-err}.

% References %
\bibliographystyle{JHEP}
\bibliography{cite}

\end{document}